\documentclass[11pt]{article}
\usepackage{amssymb,amsmath,amsfonts}
\usepackage{graphicx}
\usepackage{graphics}
\usepackage{eepic,epsfig}

1.60cm \makeatletter \@addtoreset{equation}{section}

\makeatother

\begin{document}

\title{Fermionic vacuum polarization induced  by a non-Abelian vortex}
\author{ E. R. Bezerra de Mello\thanks
	{E-mail: emello@fisica.ufpb.br} and  H. F. Santana Mota\thanks{E-mail: hmota@fisica.ufpb.br},\\
	\\
	\textit{Departamento de F\'{\i}sica, Universidade Federal da Para\'{\i}ba}\\
	\textit{58.059-970, Caixa Postal 5.008, Jo\~{a}o Pessoa, PB, Brazil}\vspace{%
		0.3cm}\\}
\maketitle

\begin{abstract}
In this paper, we analyze the fermionic condensate (FC) and the vacuum expectation
value (VEV) of the energy-momentum tensor associated with an isospin-$1/2$ charged  massive
fermionic field induced by the presence of a $SU(2)$  vortex, 
taking into account the effect of the conical geometry produced by this
object. We consider the vortex as an idealized topological defect, i.e.,
very thin, straight and carrying a magnetic flux running along its core. 
Besides the direct coupling of the fermionic field with the iso-vector gauge field, 
we also admit  the coupling  with  the scalar 
sector of the non-Abelian vortex system, expressed as a  vector in the 
three-dimensional isospace. Due to this interaction,
the FC is expressed as the sum of two contributions
associated with the two different effective masses for the $\pm 1/2$ fermionic components of 
the isospin operator, $\tau^3/2$. The VEV of the energy-tensor also presents
a similar structure. The vacuum energy density is equal 
to the radial and axial stresses. As to the azimuthal one, it is
expressed in terms of the radial derivative of energy-density. 
Regarding to the magnetic flux, both, the FC 
and the VEV of the energy-momentum tensor, can be positive or negative. 
Another interesting consequence of the interaction with the bosonic sector,  the
FC and VEV of the energy-momentum tensor, present different intensity 
for different values of the ratio between the scalar coupling constant and 
the mass of the fermionic field.
This is a new feature that the system presents.

\end{abstract}

\bigskip


\bigskip

\section{Introduction}
According to the Big Bang theory, the Universe has been undergone a series of phase 
transitions during its expansion, characterized by spontaneous symmetry 
breaking \cite{Kibble1976}. As consequence of these transitions, topological defects
 such as domain  walls, monopoles, vortices, among others  have been formed \cite{Vilenkin-Shellard}.
One of the first theoretical model  analyzing the formation of topological objects, has been
proposed by Nielsen and Olesen \cite{N-O} many years ago. They pointed out
that, in analogy with Landau-Ginzburg theory of superconductivity, the Lagrangian 
comprised by vector and  Higgs fields, that undergo spontaneously $U(1)$  
gauge symmetry breaking, presents string-like solutions,  named by  vortex. 
The influence of this object on the geometry of the spacetime
has been analyzed in \cite{DG, Laguna}, by coupling the energy-momentum tensor associated with
the Nielsen and Olesen vortex system with the Einstein equations.  
The authors showed by numerical analysis that, asymptotically, the spacetime around
the string is a Minkowski one minus a wedge. The core of the vortex has a
nonzero thickness and a magnetic flux running along it. Two years later, Linet
\cite{Linet1} was able to obtain exact solutions for the complete system,
considering specific values for the parameters.  Linet showed that the
structure of the two-geometry orthogonal to the string is a conical one, with the
conicity parameter being expressed in terms of the energy per unit length of
the vortex. In addition to the Abelian vortex, in \cite{N-O}  has also been analyzed 
the formation of  a $SU(2)$ vortex, where it has been proved that  it is necessary 
the presence of two bosonic iso-vectors coupled to the gauge field.

More recently, the analysis of the geometry of the spacetime due to the
presence of the $SU(2)$ non-Abelian Higgs
vortex, has been developed in \cite{Padua_2015}. There it was shown that, similarly 
what happens in the Abelian system, static and cylindrically solution of the system is present,
and asymptotically the space is flat with a deficit planar angle. In fact
for the non-Abeian vortex, the deficit planar angle in the two-geometry orthogonal 
to the string is larger than the Abelian case for the same energy scale
where the gauge symmetry is spontaneously broken. 

In this way, the vortices solutions of the Nielsen and Olesen model, can also be 
considered as good candidates to represent a cosmic sting.

The geometry produced by an idealized cosmic string, i.e., infinitely long, 
straight with zero transverse size,  has a conical topology with a planar
angle deficit in the two-geometry orthogonal to the string proportional 
to its linear mass density. The vacuum
polarization effects in quantum field theory induced by this object
have been considered in a large number of papers. Specifically the
analysis for the VEV of the energy-momentum tensor,
has been developed for scalar, fermionic and electromagnetic
fields \cite{Hell86}-\cite{BezeKh06}. For charged fields, considering the
presence of a magnetic flux running along the cosmic strings, there appear
additional contributions to the corresponding vacuum polarization effects
\cite{charged}-\cite{Spin08}.

The analysis of the dynamics of fermionic fields interacting 
with vortices in Abelian and $SU(2) $ non-Abelian 
gauge theories under relativistic quantum mechanics viewpoint, 
has been developed in \cite{Vega78}; moreover 
in a $(1+2)$-dimensions, considering quantum fluctuations around the non-Abelian vortex
fields, the fermionic condensate, induced  quantum numbers and Wilson loops, have been
investigate in \cite{Mello87}, \cite{Mello88a} and \cite{Mello88b}, respectively.

As far as we know, the analysis of fermionic vacuum polarization has been developed 
considering Abelian vortex only; therefore, the main objective of this paper is to 
investigate this effects associated with a isospin-$1/2$ charged massive fermionic field, induced by
the presence of a $SU(2)$ vortex, considering the effect of the conical
geometry produced by this object, and also taking into account the
coupling of the fermionic field with the iso-scalar sector of the system. 
Specifically we want to calculate the fermionic condensate (FC),  $\langle{\bar{\Psi}}\Psi\rangle$, and the 
vacuum expectation value (VEV) of the corresponding energy-momentum 
tensor, $\langle T^\mu_\nu\rangle$. In order to develop these calculations, we will assume 
the idealized case, where the size of the vortex is considered zero.

The paper is organized as follows. In Sec. \ref{Model} we present
the system that we want to investigate. First we introduce the $SU(2)$ 
non-Abelian vortex system, and the influence of this system on the
geometry of the space-time. It is also presented the
Lagrangian density associated with
the isospin$-1/2$ fermionic field coupled to a non-Abelian gauge fields, and 
the iso-vector bosonic sector. Also we provide 
the  complete set of normalized positive- and negative-energy
fermionic wave-functions. By using the mode-summation procedure, we furnish 
in Sec. \ref{FCond} the main steps to calculate the fermionic condensate, FC.
It is presented some analytical results for the FC for some limiting values of the
parameters of the system, and relevant plots displaying the behavior of FC. 
The calculation of the VEV of the energy-momentum tensor is developed in  
Sec. \ref{EMT0}, by using the mode-summation procedure; there we 
also provide some asymptotic expressions for the energy-density for
specific values of the parameters, and also relevant plots. The
 properties satisfied by this tensor are analyzed in Sec. \ref{sec5}.
 Finally in Sec. \ref{conc} we summarize the most relevant results of this paper, and present
 a brief comments about them.
 Throughout the paper we use the units with $G=\hbar =c=1$

\section{The model}

\label{Model}

\subsection{$SU(2)$ vortex}

The $SU(2)$ vortex configuration can be obtained by the matter Lagrangian density below:
\begin{equation}
 \mathcal{L}_m = -\frac{1}{4}F^a_{\mu \nu}F^{\mu \nu a} + \frac{1}{2}(D_{\mu}\varphi^a)^2 + 
 \frac{1}{2}(D_{\mu}\chi^a)^2 - V(\varphi^a, \chi^a),
 \quad a = 1, 2, 3  \  , \label{eqLagrangian}
\end{equation}
being the field strength tensor given by,
\begin{equation}
 F^{a}_{\mu \nu} = \partial_{\mu} A^a_{\nu} - \partial_{\nu} A^a_{\mu} + g \epsilon^{abc}A^b_{\mu}A^c_{\nu}  \  ,
\end{equation}
expressed in terms of  the $SU(2)$ gauge potential $A^b_{\mu}$, with $g$ being the gauge coupling constant. 
The covariant derivatives of the isovectors Higgs fields are given by  \cite{N-O},
\begin{equation}
 D_{\mu}\varphi^a = \partial_{\mu} \varphi^a + g \epsilon^{abc}A^b_{\mu}\varphi^c  \  ,
\end{equation}
\begin{equation}
 D_{\mu}\chi^a = \partial_{\mu} \chi^a + g \epsilon^{abc}A^b_{\mu}\chi^c  \  .
\end{equation}
The latin indexes denote the internal gauge groups $(a,b = 1,2,3)$. The analysis of the
influence of this system on the geometry of the spacetime was developed in \cite{Padua_2015} 
considering a non-negative interaction potential, $ V(\varphi^a, \chi^a)$, 
that admits spontaneous symmetry breaking.
Moreover,  the following configurations for the iso-scalar fields have been taking into account,
\begin{equation}\label{iso-vector}
	\varphi(r,\phi)= f(r)
	\left( \begin{array}{c}
		\cos(\phi)\\ 
		\sin(\phi)\\
		0
	\end{array} \right) \  , 
\end{equation}
\begin{equation}
	\chi(r,\phi)= h(r)
	\left( \begin{array}{c}
		-\sin(\phi)\\ 
		\cos(\phi)\\
		0
	\end{array} \right)  \  
\end{equation}
and for the iso-vector field,
\begin{eqnarray}
\label{iso_vector}
	A_t^a=A_r^a=A_z^a=0 \ , \ {\rm and} \ A_\phi^a=-\frac{A(r)}{g}\delta_{a,3}  \  ,   \quad  a = 1, 2, 3 \  . 
\end{eqnarray}
We can see that both iso-vectors bosonic fields satisfy the orthogonality condition,  $\varphi^a\chi^a=0$.

Considering cylindrically symmetric line element invariant under boosts along $z-$direction below, 
\begin{equation}
 ds^2 = N^2(r)dt^2 - dr^2 - L^2(r)d\phi^2 - N^2(r)dz^2 \  ,
 \label{ds}
\end{equation}
in \cite{Padua_2015}, was shown that for points outside the vortex, the metric tensor functions, 
$N(r)$, $L(r)$ and the gauge function $A(r)$, satisfy the asymptotic behavior below,
\begin{eqnarray}
\label{Ansatz_1}
N\to 1 \  ,  \ L\to\beta r \   , \ A\to 1 \   .
\end{eqnarray}
being $\beta$ a constant smaller than unity, providing a planar angle deficit, $\delta=2\pi(1-\beta)$, 
on the two-geometry orthogonal to the vortex.  

The idealized vortex configuration corresponds to the situation where all the fields and metric tensor can 
be represented by their corresponding vacuum values. This approximation is valid  in the analysis of phenomena 
where the energies considered are smaller than the energy scale where the gauge symmetry is broken.

\subsection{Fermionic field}
Now let us consider the charged massive isopin$-1/2$ fermionic 
field coupling to isovectors, $\varphi^a$, $\chi^a$ and $A_\mu^a$  
in curved spacetime. The corresponding  Lagrangian density is,
\begin{eqnarray}
	\label{Lagra}
{\cal{L}}&=&{\bar{\psi}}\left[i\gamma^\mu(\nabla_\mu+ieA_\mu^a\tau^a/2)-m\right]\psi\nonumber\\
&+&{\bar{\psi}}\tau^a\psi(g_1\varphi^a+g_2\chi^a)  \  , 
\end{eqnarray}
where $\tau^a$ are the Pauli matrices acting on the iospin indices, 
and  $\gamma ^{\mu }$  the Dirac matrices in curved spacetime. In \eqref{Lagra},
$e$, $g_1$ and $g_2$ correspond the coupling constants between the
fermionic field to the gauge and scalars iso-vectors, $A_\mu^a$, $\varphi^a$ and 
$\chi^a$, respectively\footnote{In principle there is no need that the electric charge
of a fermion coupling to the gauge field of vortex be quantized \cite{Jackiw_1981}.
So in this paper we will assume that $g$, in general, is different from $e$.  }. 

The fermionic covariant derivative
reads,
\begin{eqnarray}
\nabla_\mu=\partial_\mu+\Gamma_\mu  \  ,
\end{eqnarray}
being $ \Gamma _{\mu }$  the spin connection. Both matrices are given in terms of
the flat spacetime Dirac matrices, $\gamma ^{(a)}$, by the relations,
\begin{equation}
\gamma ^{\mu }=e_{(a)}^{\mu }\gamma ^{(a)}\ ,\ \Gamma _{\mu }=\frac{1}{4}%
\gamma ^{(a)}\gamma ^{(b)}e_{(a)}^{\nu }e_{(b)\nu ;\mu }\ .  \label{Gammamu}
\end{equation}%
In (\ref{Gammamu}), $e_{(a)}^{\mu }$ represents the tetrad basis satisfying
the relation $e_{(a)}^{\mu }e_{(b)}^{\nu }\eta ^{ab}=g^{\mu \nu }$, with $
\eta ^{ab}$ being the Minkowski spacetime metric tensor.

The analysis that we want to  perform here is the calculations of the fermionic vacuum fluctuations
induced by the presence of the vortex configuration. To make this analysis simpler 
we will assume the idealized configuration for the vortex. In this way we 
consider that the geometry of the spacetime can be written, by using cylindrical coordinates,
through the line element below:
\begin{equation}
ds^{2}=dt^{2}-dr^{2}-r^{2}d\phi ^{2}-dz^{2}  \   ,  \label{ds21}
\end{equation}
where the coordinates take values in the ranges $r\geqslant 0$, 
$\phi\in[0, \ 2\pi/q]$, $-\infty <(t, \  z)<+\infty $. In the metric tensor above, 
we have redefined  the azimuthal angular coordinate incorporating the parameter $\beta$ 
specified in \eqref{Ansatz_1}; so the parameter $q=1/\beta\geq 1$.
 In the geometry described by (\ref{ds21})
the gamma matrices can be taken in the form \cite{Saha13}-\cite{Saha14}:
\begin{equation}
\gamma ^{0}=\gamma ^{(0)}=\left(
\begin{array}{cc}
1 & 0 \\
0 & -1%
\end{array}%
\right) ,\;\gamma ^{l}=\left(
\begin{array}{cc}
0 & \sigma ^{l} \\
-\sigma ^{l} & 0%
\end{array}%
\right) \ ,  \label{gamcurved}
\end{equation}%
where for the $2\times 2$   Pauli matrices $\sigma ^{l}$ in this conical geometry, 
with $l=(r,\ \phi ,\ z)$, reads,
\begin{equation}
\sigma ^{r}=\left(
\begin{array}{cc}
0 & e^{-iq\phi } \\
e^{iq\phi } & 0%
\end{array}%
\right) \ ,\ \sigma ^{\phi }=-\frac{i}{r}\left(
\begin{array}{cc}
0 & e^{-iq\phi } \\
-e^{iq\phi } & 0%
\end{array}%
\right) \ ,\ \sigma ^{z}=\left(
\begin{array}{cc}
1 & 0 \\
0 & -1%
\end{array}%
\right) \ .  \label{betl}
\end{equation}
It is easy to check that with this choice the matrices (\ref{gamcurved})
obey the Clifford algebra with the metric tensor from (\ref{ds21}).
For the spin connection and the combination appearing in the Dirac equation
we find
\begin{equation}
\Gamma _{\mu }=\frac{1-q}{2}\gamma ^{(1)}\gamma ^{(2)}\delta _{\mu
}^{\phi} \ , \ \gamma ^{\mu }\Gamma _{\mu }=\frac{1-q}{2r}\gamma ^{r}   \  .
\label{gammu_matrice}
\end{equation}
As to the non-Abelian gauge field we assume
\begin{eqnarray}
	A^a_\mu=(0, \ 0\, \ A^a_\phi, \ 0 )  \ .
\end{eqnarray}
The only non-vanishing component $A^3_\phi$ is related to an infinitesimally thin
magnetic flux, $\Phi$, running along the core of the vortex by
$A_\phi^3=-q\Phi/(2\pi)=-1/g$.

In the present paper we are interested in the calculation of the 
fermionic condensate, and the VEV of the energy-momentum tensor
induced by the non-Abelian vortex. For the
evaluation of these observables, a complete set of fermionic
mode-functions is needed. In order to develop this analysis, we want to exhibit below, 
the notation adopted along this paper:
\begin{eqnarray}
	\label{def_matrrix}
{\tilde{\gamma}}^\mu&\equiv&I_{(2)}\otimes \gamma^\mu=\left(
\begin{array}{cc}
\gamma^\mu & 0 \\
0 & \gamma^\mu%
\end{array}%
\right)  \ , \\
{\tilde{\tau}}^a&\equiv&\tau^a\otimes I_{(4)} \  , 
\end{eqnarray}
being $I_{(2)}$ and $I_{(4)}$ th  $2\times2$ and $4\times 4$ identities matrices.

The effective $8\times8$ matrix Hamiltonian operator associated with this system, is:
\begin{eqnarray}
\label{Hamiltonian}
\hat{H}=-i{\tilde{\alpha}}^l(\nabla_l+ieA_l^a{\tilde{\tau}}^a/2)+\beta m-\beta(g_1\varphi+g_2\chi) \   ,
\end{eqnarray}
with
\begin{eqnarray}
 \varphi=\varphi^a {\tilde{\tau}}^a \  ,  \chi=\chi^a{\tilde{\tau}}^a \  {\rm and} \ {\tilde{\alpha}}^l={\tilde{\beta}}{\tilde{\gamma}}^l  \  .
\end{eqnarray}

Due to the non-Abelian feature of the vortex solution, the system is invariant under space plus 
isospin rotations, consequently the conserved angular momentum is,
\begin{eqnarray}
\label{J_operator_1}
\hat{J}=\frac1i\frac{\partial}{\partial\phi}+\frac q2{\tilde{\Sigma}}_{(z)}+\frac q2{\tilde{\tau}}_{(3)} \  ,
\end{eqnarray}
being ${\tilde{\Sigma}}_{(3)}={\rm diag}(\sigma_3,\sigma_3,\sigma_3,\sigma_3)$, with $\sigma_3$
being the Pauli matrix. \footnote{The ${\tilde{\Sigma}}_{(z)}/2$ and ${\tilde{\tau}}_3/2 $ are the
spin and isospin generator, respectively,  of rotation around the $z-$axis in the ordinary space, and
$3-$axis in isospace, respectively.}

Using explicit matrix notation,  \eqref{J_operator_1} reads,
\begin{eqnarray}
\label{J_operator_2}
{\hat{J}}=-iI_{(8)}\partial_\phi+q{\rm diag}(1,0,1,0,0,-1,0,-1) \ .
\end{eqnarray}

Having this conserved operator, we now look for solutions of the Dirac 
equation eigenfunctions of it:
\begin{eqnarray}
{\hat{J}}\psi=qj\psi\  ,
\end{eqnarray}
with eigenvalues $j=0,\pm1, \ \pm2, \  ...  \  .$

Although in the last subsection we have briefly introduced the $SU(2)$ vortex system
and also mentioned the influence of the system on the spacetime, in the analysis
that we are going to develop in this paper, we will adopt  another direction in the isospace  
for  $\chi(r,\phi)$, given below,
\begin{equation}
\label{chi_1}
	\chi(r)= h(r)
	\left( \begin{array}{c}
		0\\ 
		0\\
		1
	\end{array} \right)  \  
\end{equation}
that preserves the condition $\varphi^a\chi^a=0$. This is  very convenient 
for our purpose\footnote{In fact this choice was presented in \cite{Vega_86} 
analyzing the electrically charged non-Abelian vortex  in $(1+2)-$dimensions 
in the presence of Chern-Simons term.}. Moreover, we will adopt in \eqref{Hamiltonian}
the coupling constant $g_1=0$. 
In this case the interaction term between the fermionic
field with the iso-scalar takes a diagonal form, and we can express the full Dirac equation,
\begin{eqnarray}
\label{Dirac_1}
\hat{H}\psi=i\partial_t\psi  \  .
\end{eqnarray}
Writing 
\begin{eqnarray}
\label{Repres}
\hat{H}=\left(
\begin{array}{cc}
H_{(+)} & 0 \\
0 & H_{(-)}%
\end{array}%
\right)  \  ,   \
\psi=\left(
\begin{array}{c}
\psi_{(+)}\\
\psi_{(-)}
\end{array}
\right) \  ,
\end{eqnarray}
two independents $4\times4$ matrices Dirac equations,
\begin{eqnarray}
\label{Dirac_2}
H_{(\pm)}\psi_{(\pm)}=i\partial_t\psi_{(\pm)}   \  ,
\end{eqnarray}
with
\begin{eqnarray}
	\label{Hamilt0}
H_{(\pm)}=-i\alpha^l(\partial_l\pm ieA_l/2)-i\frac{(1-q)}{2r}\alpha^r+\beta(m\pm g_Y)   \  .
\end{eqnarray}
In above equation $g_Y=g_2h_0$, being $h_0$ the asymptotic value 
for the  function $h(r)$, and $A_l=(0, \ A_\phi^3,  \ 0)$.  
In \eqref{Repres} the positive/negative signal in the iso-spinor refers 
to its projections, $\pm  1/2$, along $I_3=\tau^3/2$ operator. In this present analysis it is
considered only the asymptotic values of the profiles for the gauge and the scalar fields. So in this way we set 
$A_\phi^3=-q\Phi/(2\pi)$ and $h(r)= h_0$ in the Hamiltonian above.  The physical reason for this procedure
is because we are considering the fermions far from the vortex center, i.e., its
corresponding energy is much smaller than the energy  
scale where the gauge symmetry is broken. Of course a more precise analysis should take into account the
inner structure of the metric tensor, and also the gauge and scalar fields. In this way
we can say that this paper is the first step in the direction to
investigate the quantum system, fermions and non-Abelian vortex system, in a more complete treatment.\footnote{A
similar procedure where only asymptotic values for the bosonic fields were considered was in the
analysis of the catalysis of the proton decay by magnetic monopole \cite{Rubakov}.}

In order to solve the above system, we will adopt the following procedure to find positive 
energy solutions for the system:
\begin{eqnarray}
	\label{Energy_function}
H_{(\pm)}\psi_{(\pm)}({\vec{r}},t)=i\partial_t\psi_{(\pm)}({\vec{r}},t) \  ,  
\ {\rm with} \  \psi_{(\pm)}({\vec{r}},t)
\equiv e^{-iE^{(\pm)}t}\psi_{(\pm)}({\vec{r}}) \  ,
\end{eqnarray}
where due to the presence of coupling constant, $g_Y$, two effective masses 
terms take place, $m^{(\pm)}=m\pm g_Y$.  Due to the presence of these effective
masses and because we want to specify both components of the iso-spinor by the same 
set of quantum numbers, we  assumed different energies for
each component of the fermionic field.\footnote{Due to the presence of $g_Y$,
the wave-function \eqref{Repres} is not an eigenstate of the
full Dirac Hamiltonian. Booth components, iso-spin up and down, present different energies.
A similar situation also happens due to the  Zeeman effect in the analysis of electron spin
precession in an homogeneous magnetic field \cite{Greiner}.}

\subsubsection{Wave-function $\psi_{(+)}$}
For  positive projection four-component spinor $\psi_{(+)}$, and decomposing it
into upper and lower components, denoted by $\psi_{\uparrow}$ and $\psi _{\downarrow}$,
respectively, we find the equations%
\begin{eqnarray}
\left( \sigma ^{l}(\partial _{l}+ieA_{l}/2)+\frac{1-q}{2r}\sigma ^{r}\right)
\psi _{\uparrow}-i\left( E^{(+)}+m^{(+)}\right) \psi _{\downarrow} &=&0\ ,  \notag \\
\left( \sigma ^{l}(\partial _{l}+ie2A_{l}/2)+\frac{1-q}{2r}\sigma ^{r}\right)
\psi _{\downarrow}-i\left( E^{(+)}-m^{(+)}\right) \psi _{\uparrow} &=&0  \  ,   \label{phixieq}
\end{eqnarray}
with $m^{(+)}=m+ g_Y$.
Substituting the function $\psi _{\downarrow}$ from the first equation into the
second one, we obtain the second order differential equation for the spinor $%
\psi _{\uparrow}$:
\begin{equation}
\left[ \partial _{r}^{2}+\frac{1}{r}\partial _{r}+\frac{1}{r^{2}}\left(
\partial _{\phi }+ieA^3_{\phi}/2-i\frac{1-q}{2}\sigma ^{z}\right) ^{2}+
\partial _{z} ^{2}+(E^{(+)})^{2}-(m^{(+)})^{2}\right] \psi _{\uparrow}=0\ .
\label{phieq}
\end{equation}
The same equation is obtained for the spinor $\psi _{\downarrow}$.

In order to look for solution for (\ref{phieq}), we use the the ansatz
below, compatible with the cylindrical symmetry of the physical system:
\begin{equation}
\psi _{\uparrow}=e^{i(qj\phi+kz)}\left(
\begin{array}{c}
C_{1}R_{1}(r)e^{-iq\phi } \\
C_{2}R_{2}(r) 
\end{array}
\right) \  ,  \label{psi+}
\end{equation}
with $C_{1}$ and $C_{2}$ being two arbitrary constants. Substituting this function into (\ref{phieq}), we
can see that the solutions of the equations for the radial functions
regular at $r=0$, are expressed in terms of the Bessel function of the
first kind \cite{Grad}, 
\begin{eqnarray}R_{l}(r)=J_{|\nu _{l}|}(\lambda r) \ , \ {\rm for} \ l=1, \ 2 \ ,
\end{eqnarray}
with corresponding order,
\begin{equation}
\nu _{1}=q\left( j+\alpha-1/2\right) -1/2\ ,\ \nu _{2}=q\left(
j+\alpha-1/2\right) +1/2\  .  \label{nu12}
\end{equation}
In the above expression we have introduced the notation, 
\begin{eqnarray}
	\label{Gauge}
eA^3_\phi=2q\alpha  \   .
\end{eqnarray} 
We can see that $|\nu_2|=|\nu_1|+\epsilon_{\nu_1}$, where $\epsilon_{\nu_1}$ is equal to $+1$
for $\nu_1\geq 0$ and $-1$ for $\nu_1<0$.

Having obtained $\psi_\uparrow$, we can find $\psi_\downarrow$, 
by using \eqref{phixieq}. The two-component  spinor has the form,
\begin{equation}
\psi _{\downarrow}=e^{i(qj\phi+kz)}\left(
\begin{array}{c}
B_{1}R_{1}(r)e^{-iq\phi } \\
B_{2}R_{2}(r) 
\end{array}
\right) \  .  \label{psi-}
\end{equation}
The coefficients $B_{1}$ and $B_{2}$ are related to $C_{1}$ and $C_{2}$ through \eqref{psi+},
by,
\begin{eqnarray}
B_{1} &=&\frac{1}{E^{(+)}+m^{+}}\left( C_{1}{k}-iC_{2}\epsilon _{\nu
_{1}}\lambda \right) \ ,  \notag \\
B_{2} &=&\frac{1}{E^{(+)}+m^{+}}\left( iC_{1}\lambda \epsilon _{\nu _{1}}-C_{2}{k
}\right) \ .  \label{B12}
\end{eqnarray}

The fermionic wave-function presents two 
independent coefficients. The normalization condition provides
an extra restriction, consequently one of the coefficient 
remains arbitrary. In order to determine this coefficient some additional condition
should be imposed on the coefficients. The necessity for this imposition is
related to the fact that the quantum numbers $(\lambda ,k,j)$ do not specify
the fermionic wave-function uniquely and some additional quantum number is
required. 
In order to specify the second constant we impose the condition
\begin{equation}
C_{1}/B_{1}=-C_{2}/B_{2}\ .  \label{AdCond}
\end{equation}%
By taking into account (\ref{B12}), we can write,
\begin{equation}
C_{2}=sC_{1}\ ,\ B_{1}=-sB_{2}=\frac{k-is\epsilon _{\nu
_{1}}\lambda }{E^{(+)}+m^{(+)}}C_{1}\ ,\ s=\pm 1\ .
\end{equation}
With the condition (\ref{AdCond}), the fermionic mode functions are specified by the set
of quantum numbers $\sigma =(\lambda,k,j,s)$. Of course, instead of (\ref{AdCond}) we could use another
condition. The only restriction is that the resulting  fermionic mode functions should
form a complete set. For example, the condition similar to (\ref{AdCond})
with the opposite sign of the right-hand side gives another set of fermionic mode
functions. The vacuum state is defined by the time dependence of the mode
functions and it does not depend on the particular choice of condition
required for the coefficients. Consequently, different conditions
would result the same fermionic condensate and the VEV of the energy-momentum tensor.

The negative-energy solution associated with this Hamiltonian, can also be constructed
following  analogous procedure. This state is also characterized by the same set
of quantum number, $\sigma$.

On the basis of all these considerations, the positive- and negative-energy fermionic wave
functions can be writing in a compact notation below:
\begin{eqnarray}
\label{Solution_1}
\psi _{\sigma(+) }^{(\pm)}(x)=C_{\sigma(+) }^{(\pm)}e^{\mp iE^{(+)}t+i(kz+qj\phi)
}\left(
\begin{array}{c}
J_{\beta _{j}}(\lambda r) e^{-iq\phi}\\
sJ_{\beta _{j}+\epsilon _{j}}(\lambda r) \\
 \pm\frac{{k}-is\epsilon _{j}\lambda }{E^{(+)}\pm m^{(+)}}J_{\beta _{j}}(\lambda r)e^{-iq\phi} \\
\mp s\frac{{k}-is\lambda \epsilon _{j}}{E^{(+)}\pm m^{(+)}}J_{\beta
_{j}+\epsilon _{j}}(\lambda r)
\end{array}%
\right) \    ,  
\end{eqnarray}%
with
\begin{equation}
\beta _{j}=q|j+\alpha-1/2 |-\epsilon _{j}/2 \   ,   \label{betaj}
\end{equation}%
being $\epsilon _{j}=\mathrm{sgn}(j+\alpha -1/2)$, and $C_{\sigma(+) }^{(\pm)}$ 
a normalization constant.

The energy it is expressed in terms of $\lambda$ and $k$ by the relation,
\begin{eqnarray}
	\label{Energy+}
	E^{(+)}=\sqrt{\lambda^2+k^2+(m^{(+)})^2} \   .
\end{eqnarray}

\subsubsection{Wave-function $\psi_{(-)}$}
To obtain the four-component spinor $\psi_{(-)}$, we have to adopt a similar procedure as before. 
Only small modifications on the main differential equations are required. These 
modifications are related to the sign of the charge, $e\to -e$, the definition of the
effective mass, now being $m^{(-)}=m-g_Y$, and the energy,
\begin{eqnarray}
\label{Energy-}
E^{(-)}=\sqrt{\lambda^2+k^2+(m^{(-)})^2} \   .
\end{eqnarray}
The positive and negative-energy fermionic wave functions are expressed as,
\begin{equation}
\label{Solution_2}
\psi _{\sigma(-) }^{(\pm)}(x)=C_{\sigma(-) }^{(\pm)}e^{\mp iE^{(-)}t+i(kz+qj\phi)
}\left(
\begin{array}{c}
J_{{\tilde{\beta}} _{j}}(\lambda r)\\
sJ_{{\tilde{\beta}} _{j}+{\tilde{\epsilon}} _{j}}(\lambda r)  e^{iq\phi}\\
\pm \frac{{k}-is{\tilde{\epsilon}} _{j}\lambda }{E^{(-)}\pm m^{(-)}}J_{{\tilde{\beta}} _{j}}(\lambda r) \\
\mp s\frac{{k}-is\lambda{\tilde{ \epsilon}} _{j}}{E^{(-)}\pm m^{(-)}}J_{{\tilde{\beta}}
_{j}+{\tilde{\epsilon}} _{j}}(\lambda r) e^{iq\phi}
\end{array}%
\right) \    ,  
\end{equation}%
with
\begin{equation}
{\tilde{\beta}} _{j}=q|j-\alpha+1/2 |-{\tilde{\epsilon}} _{j}/2 \   ,   \label{betaj1}
\end{equation}%
being ${\tilde{\epsilon}} _{j}=\mathrm{sgn}(j-\alpha +1/2)$, and $C_{\sigma(-) }^{(\pm)}$ 
a normalization constant.

\subsubsection{Normalization constant}

The normalization constants in (\ref{Solution_1}) and \eqref{Solution_2} can be determined from
the orthonormalization condition
\begin{equation}
\int d^{3}x\sqrt{\gamma }\ (\psi _{\sigma }^{(\pm)})^{\dagger }\psi _{\sigma'
}^{(\pm)}=\delta _{\sigma \sigma ^{\prime }}\    ,    \label{normcond}  
\end{equation}
where we assume,
\begin{eqnarray}
\label{psi_total}
\psi_\sigma^{(\pm)}=\left(
\begin{array}{c}
\psi_{\sigma(+)}^{(\pm)}\\
\psi_{\sigma(-)}^{(\pm)}
\end{array}
\right) \  ,
\end{eqnarray}
a $8-$component spinor. In  \eqref{normcond} ,  $\gamma $ represents the 
determinant of the spatial metric tensor. The delta
symbol on the right-hand side is understood as the Dirac delta function for
continuous quantum numbers, $(\lambda, \ k)$, the Kronecker delta for discrete
ones, $(j  , \  s)$. 

Considering first positive-energy solution, we get the following result,
\begin{eqnarray}
\label{Norm_1}
|C_{\sigma(+) }^{(+)}|^2\frac{16\pi^2E^{(+)}}{q\lambda(E^{(+)}+m^+)}+
|C_{\sigma(-) }^{(+)}|^2\frac{16\pi^2E^{(-)}}{q\lambda(E^{(-)}+m^-)}=1  \   .
\end{eqnarray}

Here we adopt the specific choice below for each constant, which contains
information specifically of the corresponding wave-function:
\begin{eqnarray}
\label{Const+}
|C_{\sigma(\pm) }^{(+)}|^2=\frac{q\lambda(E^{(\pm)}+m^{(\pm)})}{32\pi^2 E^{(\pm)}} \   . 
\end{eqnarray}

Adopting similar procedure, the normalization constant associated with negative-energy 
fermionic mode, is
\begin{eqnarray}
\label{Const-}
|C_{\sigma(\pm) }^{(-)}|^2=\frac{q\lambda(E^{(\pm)}-m^{(\pm)})}{32\pi^2 E^{(\pm)}} \   . 
\end{eqnarray}

\section{Fermionic condensate}
\label{FCond}

The phenomenon of Fermi condensation is important in both quantum field theories and in 
condensed matter physics. It plays a relevant role in the investigation
of superconductivity and phase transitions in models of dynamical mass 
generation and symmetry breaking. Its characteristic feature is the appearance of
non vanishing  fermion condensate (FC) defined as the expectation value
$\langle 0|\bar{\psi}\psi |0\rangle \equiv
\langle \bar{\psi}\psi \rangle $, where $|0\rangle $ is the vacuum state and
$\bar{\psi}=\psi ^{\dagger }\gamma ^{(0)}$ is the Dirac adjoint.
Various mechanisms for the formation of the FC
have been considered in the literature. They include different kinds of
interactions of fermion fields, in particular, the Nambu-Jona-Lasinio-type
models with self-interacting fields \cite{Klim88,Eliz94}.  An interesting line
of investigation  of the Fermi condensation is the
dependence of the FC on the local topology of the background
spacetime and interaction with gauge fields. Here in this section we
are interested to evaluate the FC induced by the presence of non-Abelian vortex. 

Expanding the field operator in terms of the complete set of positive- and negative-energy
solutions,  and using the anticommutation relations for the creation and
annihilation operators,  the FC can be evaluated by the following mode sum formula:
\begin{equation}
\langle \bar{\psi}\psi \rangle =-\frac{1}{2}\sum_{\sigma }\sum_{\chi
=-,+}\chi \bar{\psi}_{\sigma }^{(\chi )}\psi _{\sigma }^{(\chi )} \ .
\label{FC}
\end{equation}
In the above expression, $\psi_\sigma^{(\chi)}$, for $\chi=+1, \ -1$, represents positive
and negative energy solutions, given in \eqref{Solution_1} aand \eqref{Solution_2}. The
summation goes over the complete set of quantum numbers,
\begin{equation}
\sum_{\sigma }=\sum_{j}\int_{0}^{\infty }d\lambda \int_{-\infty }^{\infty }
dk\sum_{s =\pm 1}   \   .  \label{Sumsig}
\end{equation}%
In \eqref{Sumsig} we use the notation $\sum_{j}=\sum_{j=0,  \pm 1,\pm 2,\cdots }$. 
In addition we assume that the parameter $\alpha$ in \eqref{Gauge} is in the range,
\begin{eqnarray}
\alpha\in[-1/2, 1/2] \  .
\end{eqnarray}
In fact if we have written $\alpha$ as an integer number plus a fractional part,
$\alpha=n+\epsilon$, $n$ can be absorbed in a redefinition
of the quantum number $j$, present in the summation over this 
number, and all the physical result will depend only on the fractional 
part. This corresponds to the Aharanov-Bohm effect.

 By using the expression \eqref{psi_total}, we can see that,
\begin{eqnarray}
	{\bar{\psi}}_\sigma^{(\chi)}\psi_\sigma^{(\chi)}={\psi_\sigma^{(\chi)}}^\dagger{\tilde{\gamma}^{(0)}}
	\psi_\sigma^{(\chi)}={\psi_{\sigma(+)}^{(\chi)}}^\dagger\gamma^{(0)}\psi_{\sigma(+)}^{(\chi)}+
	{\psi_{\sigma(-)}^{(\chi)}}^\dagger\gamma^{(0)}\psi_{\sigma(-)}^{(\chi)}  \  ,
\end{eqnarray}
with $\chi=\pm$.

Substituting the fermionic modes, \eqref{Solution_1} and \eqref{Solution_2} into the above 
equation, we can observe that the terms with $s=1$ and $s=-1$ provide the same 
contributions. So after some intermediate steps, we obtain,
\begin{eqnarray}
	\label{FC1}
\langle \bar{\psi}\psi \rangle=-\frac{q}{8\pi^2}{\sum}^{'}_\sigma\left[
\frac{\lambda m^{(+)}}{E^{(+)}}
\left(J^2_{\beta_j}(\lambda r)+J^2_{\beta_j+\epsilon_j}(\lambda r)\right)
+\frac{\lambda m^{(-)}}{E^{(-)}}
\left(J^2_{{\tilde\beta}_j}(\lambda r)+J^2_{{\tilde\beta}_j+{\tilde\epsilon}_j}(\lambda r)\right)\right] \  . 
\end{eqnarray} 
In the above expression the notation ${\sum}^{'}_\sigma$ means that the 
summation over $s$ has already been developed.

In the second summation on the right hand side of \eqref{FC1}, we can change 
$j\to -j$; doing this we obtain
\begin{eqnarray}
	\tilde{\beta_j}\to\beta_j+\epsilon_j  \  ,   \  {\tilde\beta}_j+{\tilde\epsilon}_j\to\beta_j    \   .
\end{eqnarray}
So we get,
\begin{eqnarray}
	\label{FC2}
\langle \bar{\psi}\psi \rangle=-\frac{q}{4\pi^2}\sum_j\int_0^\infty d\lambda\lambda\int_0^\infty d k
\left(J^2_{\beta_j}(\lambda r)+J^2_{\beta_j+\epsilon_j}(\lambda r)\right)
\left(\frac{m^{(+)}}{E^{(+)}}+\frac{m^{(-)}}{E^{(-)}}\right)  \   . 
\end{eqnarray}

In order to obtain a more workable expression, we use the identity below,
\begin{equation}
\frac{1}{\sqrt{k^{2}+\lambda ^{2}+(m^{l})^{2}}}=\frac{2}{\sqrt{\pi }}%
\int_{0}^{\infty }ds\ e^{-(k^{2}+\lambda ^{2}+(m^{l})^{2})s^{2}}\ ,  \label{ident1}
\end{equation}%
with $l=+,-$. Substituting this identity into (\ref{FC2}), the integral over
variable $k$ is easily performed. As to the integral over $\lambda $, we use the integral
involving the square of the Bessel function from \cite{Grad}:
\begin{equation}
\int_{0}^{\infty }d\lambda \,\lambda e^{-s^{2}\lambda ^{2}} J_{\mu}^{2}(\lambda r)=
\frac{e^{-r^{2}/(2s^{2})}}{2s^{2}}I_{\mu}\left(r^{2}/(2s^{2})\right) \ ,  \label{Int-reg}
\end{equation}%
being $I_{\mu }(z)$ the modified Bessel
function. As a result, the FC is presented in the form%
\begin{eqnarray}
\langle \bar{\psi}\psi \rangle &=&-\frac{q}{(2\pi r)^{2}}\left[\frac{m^{(+)}}{2}\int_{0}^{\infty
}dy\ e^{-y-(m^{(+)})^{2}r^{2}/(2y)}\mathcal{J}(q,\alpha ,y)\right.\nonumber\\
&+&\left.\frac{m^{(-)}}{2}\int_{0}^{\infty
}dy\ e^{-y-(m^{(-)})^{2}r^{2}/(2y)}\mathcal{J}(q,\alpha ,y)\right]   \  , \label{FC3}
\end{eqnarray}%
where we have introduced the function
\begin{equation}
\mathcal{J}(q,\alpha ,y)=\mathcal{I}(q,\alpha ,y)+\mathcal{I}
(q,-\alpha ,y)  \  ,  \label{JCal}
\end{equation}%
with $\mathcal{I}(q,\alpha ,y)=\sum_{j}I_{\beta _{j}}(y)$ and $\mathcal{I
}(q,-\alpha ,y)=\sum_{j}I_{\beta _{j}+\epsilon _{j}}(y)$.

An integral representation for the function $\mathcal{I}(q,\alpha,y)$,
that allows us to extract the divergent part in the FC, has been derived in
\cite{Saha10}. Here we use that representation for the function (\ref{JCal})
\footnote{In fact the summation over the quantum number $j$  
	was developed assuming that this quantum number is semi-integer; however,
	 in this analysis, $j$ is a integer number. Redefining in 
	 \eqref{betaj}, $j-1/2=j'$, the resulting expression
	 coincides with the one given in \cite{Saha10}, so we can
	 proceed the the summation over $j'$.}:
\begin{eqnarray}
\mathcal{J}(q,\alpha,y) &=&\frac{2}{q}e^{y}+\frac{4}{\pi }
\int_{0}^{\infty }dx\,\frac{h(q,\alpha,x)\sinh x}{\cosh (2qx)-\cos
(q\pi )}e^{-y\cosh (2x)}  \notag \\
&&+\frac{4}{q}\sum_{k=1}^{p}(-1)^{k}\cos (\pi k/q)\cos (2\pi k\alpha
)e^{y\cos (2\pi k/q)}\ ,  \label{Sum01}
\end{eqnarray}%
where $p$ is an integer defined by $2p\leqslant q<2p+2$ and for $1\leqslant
q<2$ the last term on the right-hand side is absent. The function in the
integrand of (\ref{Sum01}) is given by the expression
\begin{eqnarray}
h(q,\alpha,x) =\sum_{\eta=\pm 1}\cos \left[ q\pi \left( 1/2+\eta\alpha \right) \right]
\sinh \left[ \left( 1-2\eta\alpha \right) qx\right]  \ .  \label{g0}
\end{eqnarray}%
Note that $\mathcal{J}(q,\alpha ,y)$ is an even function of $\alpha$.

We can observe that the first term on the right-hand side of (\ref{Sum01})
provides a contribution  independent of $\alpha$ and $q$ in \eqref{FC3}, that corresponds 
to the FC in the absence of the vortex.
This term is divergent. In order to obtain a finite and
well defined FC, we have to adopt some renormalization procedure. 
Because we are interested
here to calculate the contribution to the FC induced by the presence of
the vortex, the renormalization for $\langle \bar{\psi}\psi
\rangle$ reduces to subtract the exponential term in \eqref{Sum01}.
The other terms provide contributions to the FC due to
the magnetic flux and conical topology.
These terms are finite and do not require any renormalization procedure.
Substituting these terms into (\ref{FC3}), we get,
\begin{eqnarray}
	\label{FC4}
	\langle \bar{\psi}\psi \rangle ^{ren}&=&-\frac{2}{(2\pi r)^{2}}\sum_{l=+,-}m^{(l)}
\left[\sum_{k=1}^{p}(-1)^{k}\cos (\pi k/q)\cos (2\pi k\alpha)
\int_0^\infty dy e^{-2y\sin^2(\pi k/q)-(m^{(l)}r)^2/(2y)}\right.\nonumber\\
&+&	\left.\frac q\pi \int_{0}^{\infty }dx\,\frac{h(q,\alpha,x)\sinh x}{\cosh (2qx)-\cos
	(q\pi )}\int_0^\infty dy e^{-2y\cosh^2(x)-(m^{(l)}r)^2/(2y)}\right]  \  .
\end{eqnarray}
Using the integral representation below for the Macdonald function \cite{Grad},
\begin{eqnarray}
	\int_0^\infty dy e^{-\gamma y-\beta/(4y)}=\sqrt{\frac{\beta}{\gamma}}K_1(\sqrt{\beta\gamma})  \ , 
\end{eqnarray} 
 the final result for the renormalized FC can be written as:
\begin{eqnarray}
\langle \bar{\psi}\psi \rangle ^{\mathrm{ren}} &=&-\frac{1}{\pi ^{2}}
\sum_{l=+,-}(m^{(l)})^3\left[\sum_{k=1}^{p}(-1)^{k}\cos (\pi k/q)\cos (2\pi k\alpha
)f_{1}(2|m^{(l)}|rs_{k})\right. \notag \\
&&+\left. \frac{q}{\pi }\int_{0}^{\infty }dx\frac{h(q,\alpha,x)\sinh x}{%
\cosh (2qx)-\cos (q\pi )}f_{1}(2|m^{(l)}|r\cosh x)\right] \ ,  \label{FC4}
\end{eqnarray}
where we have adopted the notations
\begin{equation}
f_{\nu }(x)=K_{\nu }(x)/x^{\nu },\;s_{k}=\sin (\pi k/q)   \  .   \label{sk}
\end{equation} 
Because \eqref{FC4} is an odd function of $m^{(\pm)}$, the FC vanishes for massless field.

We can express Eq. \eqref{FC4} in terms of the parameter $\delta=g_Y/m$, defined as
the ratio between the coupling constant, $g_Y=g_2h_0$, defined below \eqref{Hamilt0}
and the mass. It reads,
\begin{eqnarray}
	\label{FC5}
\langle \bar{\psi}\psi \rangle ^{\mathrm{ren}}&=&-\frac{m^2}{2 \pi ^{2}r}\left\{(1+\delta)^2
\left[\sum_{k=1}^{p}(-1)^{k}\cot (\pi k/q)\cos (2\pi k\alpha)
K_{1}(2|1+\delta|mrs_{k})	\right.\right.	\nonumber\\
&+&\left.\left.\frac{q}{\pi }\int_{0}^{\infty }dx\frac{h(q,\alpha,x)\tanh x}{
	\cosh (2qx)-\cos (q\pi )}K_{1}(2|1+\delta|mr\cosh x)\right]+(\delta\to-\delta)\right\}  \ .
\end{eqnarray}

We can investigate the behavior of the above expression in the region $mr<<1$. Using the 
expression for the Macdonald function for small argument \cite{Abra}, we get,
\begin{eqnarray}
	\label{FC5m}
	\langle \bar{\psi}\psi \rangle ^{\mathrm{ren}}&\approx&-\frac{m}{4 \pi ^{2}r^2}\left\{(1+\delta)
	\left[\sum_{k=1}^{p}(-1)^{k}\frac{\cot (\pi k/q)}{\sin(\pi k/q)}\cos (2\pi k\alpha)
	\right.\right.	\nonumber\\
	&+&\left.\left.\frac{q}{\pi }\int_{0}^{\infty }dx\frac{h(q,\alpha,x)}{
		\cosh (2qx)-\cos (q\pi )}\frac{\tanh x}{\cos(x)}\right]+(\delta\to-\delta)\right\}  \ .
\end{eqnarray}
Moreover, at large distance from the string, i.e., for $mr>>1/|\delta\pm1|$ and for $q\geq2$, the dominant
contribution in the \eqref{FC5} comes from the term $k=1$. So, for the leading term we find
\begin{eqnarray}
	\label{FCAsymp}
\langle \bar{\psi}\psi \rangle ^{\mathrm{ren}}\approx \frac1{\sqrt{2}}\left(\frac{m}{2\pi r}\right)^{3/2}
\frac{\cos(\pi/q)\cos(2\pi\alpha)}{\sin^{3/2}(\pi/q)}\left[(1+\delta)^{3/2}e^{-2|1+\delta|mr\sin(\pi/q)}+(\delta\to -\delta)\right]  \ .
\end{eqnarray}
For $1\leq q < 2$ the sum over $k$ does not exist and the integral terms are
suppressed by the factor $e^{-2|1\pm \delta|mr}$.

 We also can see that for
 $\delta=0$,  \eqref{FC5} reproduces previous result \cite{Saha14}; in addition, admitting 
 $\alpha=0$, we get the expression given in \cite{Mello13}. For fixed $mr$, in the limit
 $\delta\to\pm 1$, the leading order in \eqref{FC5} is,
 \begin{eqnarray}
 	\label{FC6}
 	\langle \bar{\psi}\psi \rangle ^{\mathrm{ren}}&{\approx}&-\frac{2m^2}{\pi ^{2}r}
 	\left[\sum_{k=1}^{p}(-1)^{k}\cot (\pi k/q)\cos (2\pi k\alpha)
 	K_{1}(4mrs_{k})	\right.	\nonumber\\
 	&+&\left.\frac{q}{\pi }\int_{0}^{\infty }dx\frac{h(q,\alpha,x)\tanh x}{
 		\cosh (2qx)-\cos (q\pi )}K_{1}(4mr\cosh x)\right]  \ .
 \end{eqnarray}

\begin{figure}[h]
	\centering
	{\includegraphics[width=0.48\textwidth]{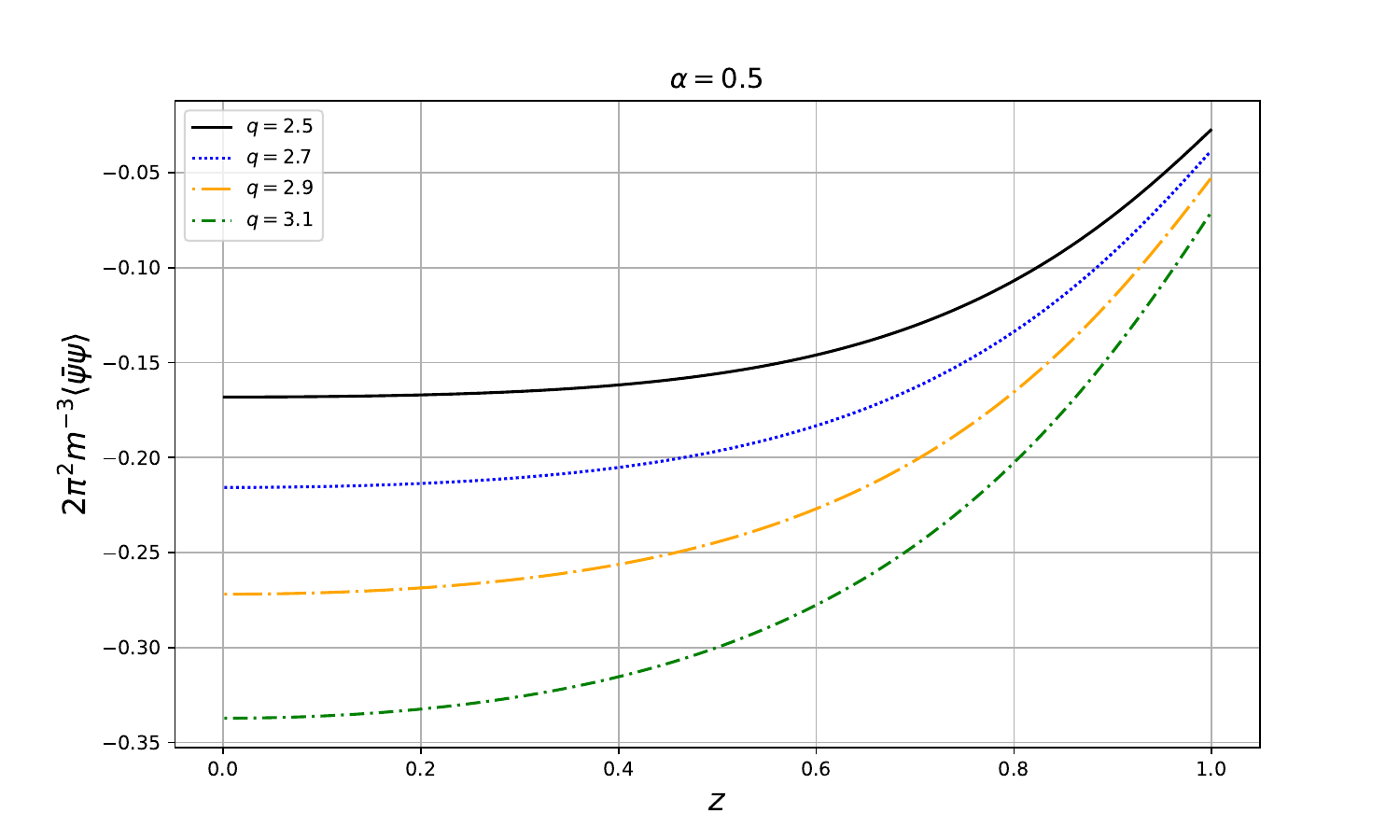}}
	\hfill
	{\includegraphics[width=0.48\textwidth]{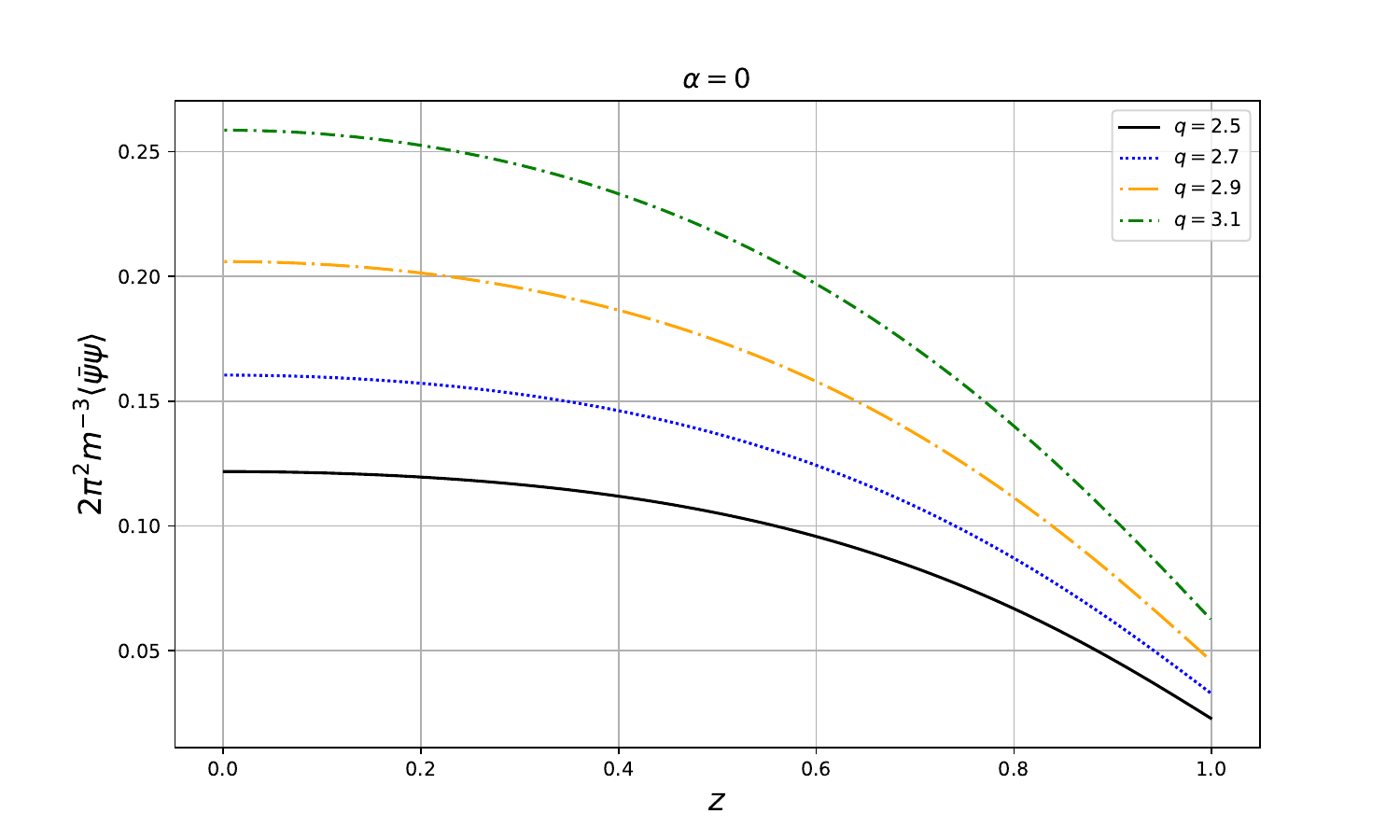}}
	\caption{The quantity $2\pi^2\langle \bar{\psi}\psi \rangle ^{\mathrm{ren}}/m^3$, is exhibited as 
		function of $\delta$ for $\alpha=1/2$  (left plot) and $\alpha=0$ (right plot)
		for  $q=2.5, \ 2.7, \ 2.9, \ 3.1  $.  In both plots we consider
		$\delta$ in the interval $[0, \ 1]$ and assume $mr=1.0$.}
	\label{fig01}
\end{figure}

In Fig. \ref{fig01}  we exhibit the behavior of the FC  as function of $\delta$ 
for $\alpha=1/2$ (left plot) and $\alpha=0$ (right plot),  considering different values of $q$.
Because \eqref{FC5} is a even function of $\delta$ 
we assume only this parameter in the interval $[0, \ 1]$. Moreover, we take
 $mr=1$.  As we can see the module of the intensity of the FC increases with $q$
 for a given value of $\delta$.

\begin{figure}[h]
	\centering
	{\includegraphics[width=0.48\textwidth]{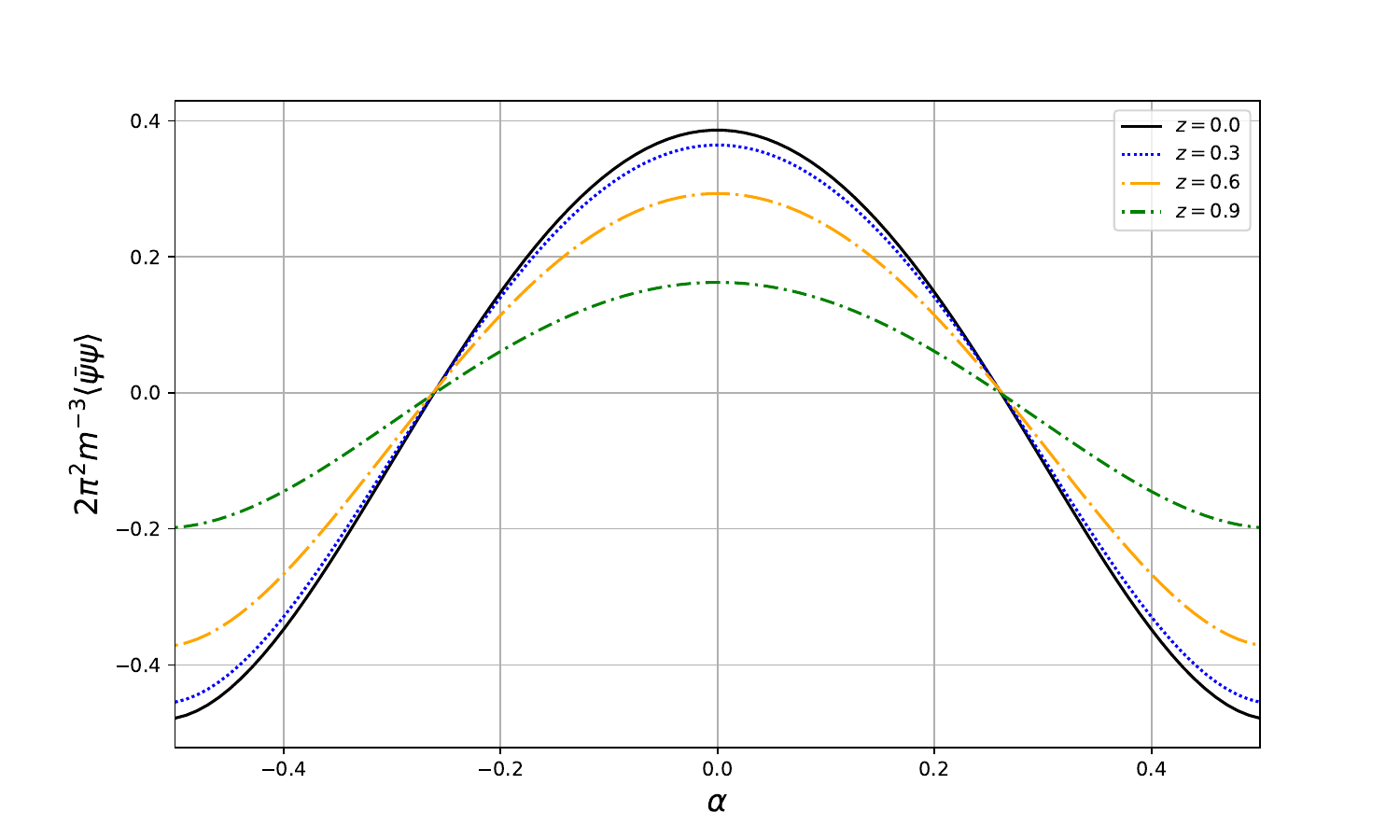}}
	\caption{The FC, $2\pi^2\langle \bar{\psi}\psi \rangle ^{\mathrm{ren}}/m^3$,
		 is exhibited as function of $\alpha$ for  
		$\delta=0.0, \ 0.3, \  0.6, \ 0.9 $. In this plot was considered 
		$mr=1$ and $q=3.5$.}
	\label{fig02}
\end{figure}

Figure \ref{fig02} presents the dependence of the FC as function of $\alpha$ 
for different values of $\delta$, cosidering $mr=1$ and $q=3.5$. By this plot we see that the intensity of 
the FC is larger for $\delta=0$ and $\alpha=0$, i.e., in absence of interaction
between the fermionic field with the scalar one, $\chi$, and the magnetic flux. The physical
explanation for this result resides in the fact that for massless field \eqref{FC5}  vanishes. 
So, for a given $\alpha$, considering $\delta\neq 0$ there is an imbalance between the contributions of both
effective masses, decreasing the magnitude of FC. 

\section{Energy-momentum tensor}

\label{EMT0}
The analysis of VEV of the energy-momentum tensor is one of the most relevant 
physical quantity associated with the quantum vacuum. Besides to describing the
distribution of the energy density and vacuum stresses around the non-Abelian vortex, 
it appears as a source of gravity in the semiclassical Einstein
equations and provides the backreaction of quantum effects on the
gravitational field.

So, motived by these facts, in this section we want to calculate the VEV of the energy-momentum tensor. 
For a charged field coupled with an electromagnetic field and in curved space,
this quantity can be calculated using the expression below, by expanding the fermionic
operator in terms of the complete set of positive and negative normalized wave-functions:
\begin{equation}
	\left\langle 0|T_{\mu \nu }|0\right\rangle =\left\langle T_{\mu \nu
	}\right\rangle =-\frac{i}{4}\sum_{\sigma }\sum_{\chi =-,+}\chi {\left[ \bar{%
			\psi}_{\sigma }^{(\chi )}{\tilde{\gamma}} _{(\mu }\mathcal{D}_{\nu )}\psi _{\sigma
		}^{(\chi )}-(\mathcal{D}_{(\mu }\bar{\psi}_{\sigma }^{(\chi )}){\tilde{\gamma}} _{\nu
			)}\psi _{\sigma }^{(\chi )}\right] }\   ,  \label{EMT}
\end{equation}
where $\mathcal{D}_{\mu }{\bar{\psi}}=\partial _{\mu }{\bar{\psi}}-ieA_{\mu }%
{\bar{\psi}}-{\bar{\psi}}\Gamma _{\mu }$, and the brackets in the index
expression means the symmetrization over the enclosed indexes.

\subsection{Energy density}

Let us first consider the energy density, $\langle T_{0}^{0}\rangle $. By
taking into account that $A_{0}=\Gamma _{0}=0$,  we have,
\begin{eqnarray}
 \bar{\psi}_{\sigma }^{(\chi )}{\tilde{\gamma}} _{0}\mathcal{D}_0\psi _{\sigma
}^{(\chi )}= {{\psi}_{\sigma(+) }^{(\chi )}}^\dagger\partial_0\psi _{\sigma(+)}^{(\chi )}+	
{{\psi}_{\sigma(-) }^{(\chi )}}^\dagger\partial_0\psi _{\sigma(-)}^{(\chi )}  \  .
\end{eqnarray}
Now writing in a compact form $\partial_{0}\psi _{\sigma(\pm) }^{(\chi)}=
-i\chi E^{(\pm)}\psi _{\sigma(\pm) }^{(\chi)} $,
and taking explicitly the fermionic wave-functions, \eqref{Solution_1} and \eqref{Solution_2},
with the corresponding normalization constant, \eqref{Const+} and \eqref{Const-},
we can see that the contributions of the terms $s=1$ and $s=-1$ are the same. After
some intermediate steps, we get:
\begin{eqnarray}
\label{T00}
\langle T_{0}^{0}\rangle&=&-\frac{q}{4\pi^2}\sum_{l=+,-}\sum_j\int_0^\infty d\lambda \lambda
\int_0^\infty dk E^{(l)}\left(
J_{\beta_{j}}^{2}(\lambda r)+J_{\beta _{j}+\epsilon _{j}}^{2}(\lambda r)\right) \  , 
\end{eqnarray}
where we have changed $j\to-j$ in the indexes of Bessel functions associated with
the projections $-1/2$ of the  iso-spinor.

In order to present the above result in a more suitable expression for the renormalization,
we will use the identity below,
\begin{equation}
	\sqrt{k^{2}+\lambda ^{2}+(m^{(l)})^{2}}=-\frac{2}{\sqrt{\pi }}\int_{0}^{\infty
	}ds\,\partial _{s^{2}}e^{-(k^{2}+\lambda ^{2}+(m^{(l)})^{2})s^{2}}\ .  \label{Ident1}
\end{equation}
Substituting \eqref{Ident1} into \eqref{T00}, the integration over the variable $k$ 
can be developed trivially. As to the integral over $\lambda$ we use Eq. \eqref{Int-reg}. So after some 
intermediate steps, we can write,
\begin{eqnarray}
\langle T_{0}^{0}\rangle=-\frac{q}{4\pi ^{2}r^{4}}\sum_{l=+,-}
\int_{0}^{\infty }dy\
y^{1/2}\partial _{y}[y^{3/2}e^{-y-(m^{(l)})^{2}r^{2}/(2y)}\mathcal{J}(q,\alpha,y)]\ ,  \label{T001}
\end{eqnarray}
where $\mathcal{J}(q,\alpha,y)$ is explicitly given in \eqref{Sum01}. The exponential term
in that expression, $\frac2qe^y$, provides a divergent result, and corresponds to the 
contribution associated with the Minkowski background in absence of magnetic flux. 
To obtain a finite and well defined value for the energy-density we
should adopt some renormalization procedure. As we did in the FC analysis we will discard 
the exponential  term, that is equivalent to the subtraction of the Minkowskian counterterms.  
In this way, we may now integrate by parts to evaluate $\langle T_{0}^{0}\rangle^{(ren)}$:
\begin{eqnarray}
	\label{T002}
	\langle T_{0}^{0}\rangle^{(ren)}&=&-\frac{1}{2\pi^2r^4}\sum_{l=+,-}\int_0^\infty dy y 
e^{-y-(m^{(l)})^{2}r^{2}/(2y)}\left[\sum_{k=1}^{p}(-1)^{k}\cos (\pi k/q)\cos (2\pi k\alpha
)e^{y\cos (2\pi k/q)}\right.	\nonumber\\
&+&\frac q\pi\left.\int_{0}^{\infty }dx\,\frac{h(q,\alpha,x)\sinh x}{\cosh (2qx)-\cos
	(q\pi )}e^{-y\cosh (2x)}\right]  \  .
\end{eqnarray}
After some minor simplifications, and using the integral representation for the 
the Macdonald function, $K_2(z)$, below \cite{Grad},
\begin{eqnarray}
K_2(z)=\frac 2{z^2}\int_0^\infty dt e^{-t-z^2/(4t)} \  , 
\end{eqnarray}
we get:
\begin{eqnarray}
		\label{T003}
	\langle T_{0}^{0}\rangle^{\mathrm{ren}} &=&\frac{1}{\pi ^{2}}\sum_{l=+,-}(m^{(l)})^4\left[
	\sum_{k=1}^{p}(-1)^{k}{\cos }(\pi k/q){\cos }(2\pi k\alpha
	)f_{2}(2|m^{(l)}|rs_{k})\right.  \notag \\
	&+&\left. \frac{q}{\pi }\int_{0}^{\infty }dx\frac{h(q,\alpha ,x)\sinh x}{%
		\cosh (2qx)-\cos (q\pi )}f_{2}(2|m^{(l)}|r\cosh x)\right] \ ,  
\end{eqnarray}
with the notation \eqref{sk}.
For massless field, the above expression reduces to,
\begin{eqnarray}
	\label{T003_m}
	\langle T_{0}^{0}\rangle^{\mathrm{ren}} &=&\frac{2}{\pi ^{2}}(g_Y)^4\left[
	\sum_{k=1}^{p}(-1)^{k}{\cos }(\pi k/q){\cos }(2\pi k\alpha
	)f_{2}(2|g_Y|rs_{k})\right.  \notag \\
	&+&\left. \frac{q}{\pi }\int_{0}^{\infty }dx\frac{h(q,\alpha ,x)\sinh x}{%
		\cosh (2qx)-\cos (q\pi )}f_{2}(2|g_Y|r\cosh x)\right] \ . 
\end{eqnarray}
Finally considering the limit $r|g_Y|\to 0$, we obtain,
\begin{eqnarray}
	\label{T003_m0}
	\langle T_{0}^{0}\rangle^{\mathrm{ren}} &=&\frac{1}{4\pi ^{2}r^4}\left[
	\sum_{k=1}^{p}(-1)^{k}\frac{{\cos }(\pi k/q)}{\sin^4(\pi k/q)}{\cos }(2\pi k\alpha
	)\right.  \notag \\
	&+&\left. \frac{q}{\pi }\int_{0}^{\infty }dx\frac{h(q,\alpha ,x)}{
		\cosh (2qx)-\cos (q\pi )}\frac{\sinh x}{\cosh^4(x)}\right] \ . 
\end{eqnarray}

\subsection{Radial stress}

Our objective in this subsection is to evaluate the radial stress, $\langle
T_{r}^{r}\rangle $. In order to do that, we take in the covariant derivative
of the fermionic field $A_{r}=\Gamma _{r}=0$.
In this way, we can write,
\begin{equation}
\langle T_{r}^{r}\rangle =-\frac{i}{4}\sum_{\sigma }\sum_{\chi=+,-} \chi\left[ \bar{\psi}_{\sigma
}^{(\chi)}{\tilde{\gamma}} ^{r}(\partial _{r}\psi _{\sigma }^{(\chi)})-(\partial _{r}\bar{
\psi}_{\sigma }^{(\chi)}){\tilde{\gamma}} ^{r}\psi _{\sigma }^{(\chi)}\right] \ .
\label{TrrModeSum}
\end{equation}%
Taking the definition \eqref{def_matrrix} for the Dirac gamma matrices, the mode functions from (\ref{Solution_1}) 
and \eqref{Solution_2} with their respective normalizations constants,
into the above expression, after some intermediate steps, we arrive at,
\begin{equation}
\langle T_{r}^{r}\rangle =-\frac{q}{16\pi^2}\sum_{\sigma }\sum_{l=+,-}\epsilon
_{j}\lambda ^{3}[J_{\beta _{j}}^{\prime }(\lambda r)J_{\beta
_{j}+\epsilon _{j}}(\lambda r)-J_{\beta _{j}}(\lambda r)J_{\beta
_{j}+\epsilon _{j}}^{\prime }(\lambda r)]\frac1{E^{(l)}}\ ,  \label{Trr1}
\end{equation}
where the primes means derivative with respect to the argument of the
function. 

Using the recurrence relation involving the Bessel functions \cite{Abra} and 
develop the summation over $s$, \eqref{Trr1} can be written  as,
\begin{eqnarray}
\langle T_{r}^{r}\rangle =\frac{q}{8\pi^2}\sum_{l=+,-}\int_{-\infty}^\infty dk 
\int_{0}^{\infty }d\lambda \lambda ^{3}S(\lambda r)\frac1{\sqrt{k^2+\lambda^2+(m^{(l)})^2}}  \  ,
\label{Trr2}	
\end{eqnarray}
where
\begin{equation}
	S(x)=\sum_{j}\left[ J_{\beta _{j}}^{2}(x)+J_{\beta _{j}+\epsilon
		_{j}}^{2}(x)-\frac{2\beta _{j}+\epsilon _{j}}{x}J_{\beta _{j}}(x)J_{\beta
		_{j}+\epsilon _{j}}(x)\right] .  \label{S1}
\end{equation}

Using the identity  (\ref{ident1}) we can develop the integration over $k$ trivially. It remains 
to evaluate the integral over $\lambda$, as shown below: 
\begin{equation}
\langle T_{r}^{r}\rangle _{s}=\frac{q}{(2\pi )^{2}}\sum_{l=+,-}\int_{0}^{\infty }ds\
\frac{e^{-(m^{(l)})^{2}s^{2}}}{s}\int_{0}^{\infty }d\lambda \,\lambda
^{3}e^{-\lambda ^{2}s^{2}}S(\lambda r)\ .
\end{equation}
Now we have to proceed the integration,
\begin{eqnarray}
	\int_{0}^{\infty }d\lambda \,\lambda ^{3}{e^{-\lambda
			^{2}s^{2}}}S(\lambda r) &=&\sum_{j}\int_{0}^{\infty }d\lambda \,\lambda ^{3}e^{-\lambda
		^{2}s^{2}}\left[ J_{\beta _{j}}^{2}(\lambda r)+J_{\beta _{j}+\epsilon
		_{j}}^{2}(\lambda r)\right]  \notag \\
	&&-\sum_{j}\frac{2\beta _{j}+\epsilon _{j}}{r}\int_{0}^{\infty }d\lambda
	\,\lambda ^{2}e^{-\lambda ^{2}s^{2}}J_{\beta _{j}}(\lambda r)J_{\beta
		_{j}+\epsilon _{j}}(\lambda r)\ .  \label{calc1}
\end{eqnarray}
As to the firsts two integrals we can use the previous result, Eq. \eqref{Int-reg}, 
by applying the derivative $-\partial_{s^2}$. However to integrate the
third term in $S(\lambda r)$, we have to adopt a more subtle method. First 
we observe that we can write, 
\begin{equation}
	J_{\beta _{j}}(\lambda r)J_{\beta _{j}+\epsilon _{j}}(\lambda r)=\frac{1}{%
		2\lambda }\left( -\epsilon _{j}\partial _{r}+2\beta _{j}/r\right) J_{\beta
		_{j}}^{2}(\lambda r)  \   .
\end{equation}
With this result in hand, we can use again \eqref{Int-reg}. So, after some
intermediate steps, we arrive to,
\begin{equation}
	\int_{0}^{\infty }d\lambda {\lambda ^{2}}{e^{-\lambda ^{2}s^{2}}}\
	J_{\beta _{j}}(\lambda r)J_{\beta _{j}+\epsilon _{j}}(\lambda r)=\frac{%
		r\epsilon _{j}e^{-y}}{4s^{4}}[I_{\beta _{j}}(y)-I_{\beta _{j}+\epsilon
		_{j}}(y)]\ ,  \label{ident-1}
\end{equation}%
with $y=r^{2}/(2s^{2})$. To obtain a more workable result, in order to use
the identity (\ref{Sum01}),  we use the relation
\begin{equation}
	(1+2\epsilon _{j}\beta _{j})[I_{\beta _{j}}(y)-I_{\beta _{j}+\epsilon
		_{j}}(y)]=2\left( y\partial _{y}-y+1/2\right) [I_{\beta _{j}}(y)+I_{\beta
		_{j}+\epsilon _{j}}(y]\ .  \label{IdentD}
\end{equation}

After combine all the above results, we get:
\begin{equation}
	\langle T_{r}^{r}\rangle=\frac{q}{8\pi ^{2}r^{4}}\sum_{l=+,-}\int_{0}^{\infty
	}dy\,ye^{-y-(m^{(l)})^{2}r^{2}/(2y)}\mathcal{J}(q,\alpha ,y)\ .  \label{Trrcs2}
\end{equation}
As we have already mentioned, the exponential term in \eqref{Sum01} corresponds 
to contribution associated with the Minkowski spacetime in absence of 
magnetic flux. The renormalized expression for the radial stress, $	\langle T_{r}^{r}\rangle^{ren}$,
 can be obtained in a manifest form by subtracting this exponential term. Doing this
we can observe that the result obtained coincides with \eqref{T001} after we
have developed the integration by part. So we can conclude that,
\begin{equation}
	\langle T_{r}^{r}\rangle^{\mathrm{ren}}=\langle T_{0}^{0}\rangle^{\mathrm{ren}} \ .  \label{TrrT00}
\end{equation}

\subsection{Azimuthal stress}

In this subsection we will evaluation of the VEV for the 
azimuthal stress, $\langle T_{\phi
}^{\phi }\rangle $. In order to do that we have to take into 
account, $A_{\phi }=2q\alpha /e$ and
\begin{equation}
{\tilde{\Gamma }}_{\phi }=\frac{1-q}{2}{\tilde{\gamma}} ^{(1)}
{\tilde{\gamma}} ^{(2)}=-\frac{i}{2}
(1-q){\tilde{\Sigma}} ^{(3)}\ .
,  \label{Gam}
\end{equation}%
So this component reads,
\begin{equation}
\langle T_{\phi\phi }\rangle =-\frac{i}{4}\sum_{\sigma }\sum_{\chi=+,-}[\bar{\psi}
_{\sigma }^{(\chi)}{\tilde{\gamma}} _{\phi }{\mathcal{D}}_{\phi }\psi _{\sigma }^{(\chi)}
-({\mathcal{D}}_{\phi }\bar{\psi}_{\sigma }^{(\chi)}){\tilde{\gamma}} _{\phi }\psi _{\sigma
}^{(\chi)}]  \ .  \label{Tphi}
\end{equation}%
In the development of the term inside the bracket, it is convenient to
express the angular derivative in terms of the total angular momentum
operator: $\partial _{\phi }=iJ-i\frac q2({\tilde{\Sigma}} _{(3)}+{\tilde{\tau}}_3)$.
Also, we can observe that the anticommutator, $\{{\tilde{\gamma}} ^{\phi },{\tilde{\Sigma}}
^{(3)}\}$, which appears in this development, vanishes. So after some steps,
we get:
\begin{eqnarray}
\langle T_{\phi,\phi }\rangle&=&\frac q2\sum_{\sigma}\sum_\chi\left[(j+\alpha-1/2 )
\bar{\psi}_{\sigma(+) }^{(\chi)}\gamma _{\phi }\psi _{\sigma(+) }^{(\chi)} \right. \nonumber\\
&+&\left.(j-\alpha+1/2 )\bar{\psi}_{\sigma(-) }^{(\chi)}\gamma _{\phi }
\psi _{\sigma(-) }^{(\chi)}\right]  \  .  \label{Tpp}
\end{eqnarray}%
Now substituting the explicit expressions for positive- and negative-energy fermionic
wave-functions, Eq.s \eqref{Solution_1} and \eqref{Solution_2}, 
the Dirac matrix $\gamma ^{\phi }$ and developing  the summation over $s$ and $\chi$, we obtain
\begin{eqnarray}
\langle T_{\phi }^{\phi }\rangle&=&\frac{q^{2}}{2\pi^2 r}\sum_{j}\int_0^\infty
d\lambda \lambda^2\int_0^\infty dk \left[(j+\alpha-1/2 )\epsilon _{j}J_{\beta _{j}}(\lambda
r)J_{\beta _{j}+\epsilon _{j}}(\lambda r)\frac1{E^{(+)}}\right. \nonumber\\
&+&\left.(j-\alpha+1/2 ){\tilde{\epsilon}} _{j}J_{{\tilde{\beta}} _{j}}(\lambda
r)J_{{\tilde{\beta}} _{j}+{\tilde{\epsilon}} _{j}}(\lambda r)\frac1{E^{(-)}}\right]  \  .  \label{Tpp1}
\end{eqnarray}
Finally changing $j\to-j$ in the second term inside the bracket, we get a 
 more convenient expression to work:
\begin{eqnarray}
	\label{Tpp2}
	\langle T_{\phi }^{\phi }\rangle=\frac{q^{2}}{2\pi^2 r}\sum_{j}\epsilon _{j}
	(j+\alpha-1/2 )\int_0^\infty dk
\int_0^\infty 	d\lambda \lambda^2 J_{\beta _{j}}(\lambda
	r)J_{\beta _{j}+\epsilon _{j}}(\lambda r)\sum_{l=+,-}\frac1{E^{(l)}}  \  .
\end{eqnarray}
To continue our development we will use \eqref{ident1}. Doing this the integral over
$k$ is trivially performed. As to the integral over $\lambda$ we use \eqref{ident-1},
so we get,
\begin{eqnarray}
\langle T_{\phi }^{\phi }\rangle&=&\frac{q}{8\pi^2}\sum_{l=+,-}\int_0^\infty \frac{ds}{s^5}
e^{-(m^{(l)})^2s^2-r^2/(2s^2)}\sum_jq(j+\alpha-1/2)\nonumber\\
&\times& \left[I_{\beta_j}(y)-I_{\beta_j+\epsilon_j}(y)\right]_{y=r^2/(2s^2)} \  .
\label{Tpp3}
\end{eqnarray}
Using the fact that $q(j+\alpha-1/2)=\epsilon_j\beta_j+1/2$ and the identity \eqref{IdentD},
we can present the above expression by,
\begin{eqnarray}
\label{Tpp4}
\langle T_{\phi }^{\phi }\rangle=\frac{q}{8\pi^2}\sum_{l=+,-}\int_0^\infty \frac{ds}{s^5}
e^{-(m^{(l)})^2s^2}(y\partial_y+1/2)e^{-y}\mathcal{J}(q,\alpha ,y)_{y=r^2/(2s^2)} \  .	
\end{eqnarray}
Moreover, using the identity below into \eqref{Tpp4},
\begin{eqnarray}
(y\partial_y+1/2)e^{-y}\mathcal{J}(q,\alpha ,y)_{(y=r^2/(2s^2))}=1/2(r\partial_r+1)
e^{-r^2/(2s^2)}	\mathcal{J}(q,\alpha ,r^2/(2s^2))  
\end{eqnarray}	
and changing the variable of integration by, $y=r^2/(2s^2)$, we can 
easily  shown that,
\begin{eqnarray}
	\label{TppT00}
\langle T_{\phi }^{\phi }\rangle^{ren}=(r\partial_r+1)\langle T_{0 }^{0}\rangle^{ren} \  .
\end{eqnarray}

An explicit expression for $\langle T_{\phi }^{\phi }\rangle^{ren}$ can be derived 
by applying the radial differential operator on the energy density, \eqref{T003},
and using the identity,
\begin{eqnarray}
	\label{Identity_f}
\partial_x(xf_2(x))=-f_1(x)-3f_2(x)  \  .
\end{eqnarray}

\subsection{Axial stress}

In the calculation of the axial stress, we have to consider $A_{z}=\Gamma_z=0$.
So, for this case we have ${\mathcal{D}}_{z}\psi _{\sigma }^{(\chi)}=\partial_z\psi _{\sigma }^{(\chi)}=i{k}\psi _{\sigma }^{(\chi)}$.
Consequently we have,
\begin{eqnarray}
	\langle T_{z}^z\rangle=\frac12\sum_\sigma\sum_{\chi}\chi k\left[{\bar{\psi}}_{\sigma(+)}^{(\chi)}
	{\gamma}^z\psi_{\sigma(+)}^{(\chi)}+{\bar{\psi}}_{\sigma(-)}^{(\chi)}
	{\gamma}^z\psi_{\sigma(-)}^{(\chi)}	\right] \   .  \label{Tzz}
\end{eqnarray}

Substituting the expressions for positive- and negative-energy mode 
functions, Eq.s \eqref{Solution_1}
and \eqref{Solution_2}, into the above expression, develop the 
summations over $s$ and $\chi$,
and conveniently changing the quantum number $j\to-j$ in the second 
contribution inside the bracket, we get
\begin{equation}
\langle T_{z}^{z}\rangle =\frac{q}{4\pi^2}\sum_{l=+,-}\int_{0}^{\infty
}d\lambda \lambda \int_0^\infty dk k^2\sum_j[J_{\beta _{j}}^{2}(\lambda r)
+J_{\beta _{j}+\epsilon _{j}}^{2}(\lambda r)]\frac1{E^{(l)}}
\ .  \label{Tzz1}
\end{equation}
Using the identity \eqref{ident1}, and the previous integral over 
the square of Bessel functions, Eq. 
\eqref{Int-reg}, one obtains,
\begin{eqnarray}
\langle T_{z}^{z}\rangle =\frac{q}{16\pi^2}\sum_{l=+,-}\int_0^\infty 
\frac{ds}{s^5} \ e^{-(r^2/(2s^2)+(m^{(l)})^2s^2)}\mathcal{J}(q,\alpha ,r^2/(2s^2))   \   , 
\label{Tzz2}
\end{eqnarray}
where we have used the general expression for the summation over the
modified Bessel function, \eqref{JCal}. 

Finally changing the variable of integration, $y=r^2/(2s^2)$, we can 
see that,
\begin{eqnarray}
	\label{TzzT00}
	\langle T_{z}^{z}\rangle^{ren}=\langle T_{0}^{0}\rangle^{ren}  \  . 
\end{eqnarray}
The above result is consequence of the invariance of the system with respect to a boost along the $z$ coordinate.

\section{Properties of the VEV of the energy-momentum tensor}
\label{sec5}

Here we want to investigate the most important properties obeyed 
by the VEV of the energy-momentum tensor. They are the trace relation,
that that establishes the relation between its trace and the femionic condensate, 
and its conservation condition, necessary to be considered as the backreaction of the
quantum fermionic field on the Einstein equation. 

For the system under consideration, we can express,
\begin{eqnarray}
	\langle T^\mu_\nu\rangle^{ren}=\langle T^\mu_\nu\rangle^{ren(+)}
	+\langle T^\mu_\nu\rangle^{ren(-)} \  ,
\end{eqnarray}
where the positive/negative signal corresponds to the signal in the effective mass
$m^{(\pm)}=m\pm g_Y$. (See Eq.s \eqref{T003}, \eqref{TrrT00}, \eqref{TppT00}, 
\eqref{Identity_f} and	\eqref{TzzT00}). On the other hand the FC has a similar 
structure. So we can write,
\begin{eqnarray}
\langle{\bar{\psi}}\psi\rangle^{ren}=\langle{\bar{\psi}}\psi\rangle^{ren(+)}
+\langle{\bar{\psi}}\psi\rangle^{ren(-)} \  ,
\end{eqnarray}
as shown in \eqref{FC4}. By using the explicit expressions for the
components of the VEV of the energy-momentum tensor and FC, we can verify,
\begin{eqnarray}
	\langle T^\mu_\mu\rangle^{ren(\pm)}=m^{(\pm)}\langle{\bar{\psi}}\psi\rangle^{ren(\pm)} \  . 
\end{eqnarray}

 As to the conservation condition, $\nabla _{\mu}\langle 
T_{\nu }^{\mu }\rangle =0$, for the problem 
under considerations, it reduces to a single expression, 
\begin{eqnarray}
	\partial_{r}(r\langle T_{r}^{r}\rangle )=\langle T_{\phi }^{\phi }\rangle \  .
\end{eqnarray}
According to the relations \eqref{TrrT00} and \eqref{TppT00}, we conclude that the VEV of the energy-momentum tensor is conserved.

Another point that we want to explore is the behavior of the
energy-density with the variable $\delta$. To obtain this
information let us express Eq. \eqref{T003} as,
\begin{eqnarray}
	\label{T00z}
\langle T_{0}^{0}\rangle^{\mathrm{ren}} &=&\frac{m^2}{4\pi ^{2}r^2}\left\{(1+\delta)^2\left[
\sum_{k=1}^{p}(-1)^{k}\frac{{\cos }(\pi k/q)}{\sin^2(\pi k/q)}{\cos }(2\pi k\alpha
)K_{2}(2m|1+\delta|rs_{k})\right. \right. \notag \\
&+&\left. \frac{q}{\pi }\int_{0}^{\infty }dx\frac{h(q,\alpha ,x)}{
	\cosh (2qx)-\cos (q\pi )}\frac{\sinh(x)}{\cosh^2(x)}K_{2}(2m|1+\delta|r\cosh x)\right] \nonumber\\
&+&\left.(\delta\to-\delta)\right\} \ .
\end{eqnarray}

As in the FC case for $\delta=0$, the above expression reproduces previous result 
given in \cite{Saha14}. Moreover, in the limit $\delta\to\pm 1$,  we obtain 
\begin{eqnarray}
	\label{T00z1}
	\langle T_{0}^{0}\rangle^{\mathrm{ren}} &=&\frac{1}{\pi ^{2}r^2}\left\{m^2\left[
	\sum_{k=1}^{p}(-1)^{k}\frac{{\cos }(\pi k/q)}{\sin^2(\pi k/q)}{\cos }(2\pi k\alpha
	)K_{2}(4mrs_{k})\right. \right. \notag \\
	&+&\left. \frac{q}{\pi }\int_{0}^{\infty }dx\frac{h(q,\alpha ,x)}{
		\cosh (2qx)-\cos (q\pi )}\frac{\sinh(x)}{\cosh^2(x)}K_{2}(4mr\cosh x)\right] \nonumber\\
	&+&\frac1{8r^2}\left[
	\sum_{k=1}^{p}(-1)^{k}\frac{{\cos }(\pi k/q)}{\sin^4(\pi k/q)}{\cos }(2\pi k\alpha)\right.\nonumber\\
	&+&\left.\left.\frac{q}{\pi }\int_{0}^{\infty }dx\frac{h(q,\alpha ,x)}{
		\cosh (2qx)-\cos (q\pi )}\frac{\sinh(x)}{\cosh^4(x)}\right]
	\right\} \ .
\end{eqnarray}
So for $mr>>1$ the above expression is dominated by the second contribution 
presenting a $1/r^4$ decay. 

At large distance from the string and for $q\geq2$, the dominant
contribution in the \eqref{T00z} comes from the term $k=1$. Under this 
condition the dominant contribution is given by,
\begin{eqnarray}
	\langle T_{0}^{0}\rangle^{\mathrm{ren}}&{\approx}&-\frac{m^4}{8\pi^{3/2}}\frac{\cos(\pi/q)\cos(2\pi\alpha)}
	{(mr\sin(\pi/q))^{5/2}}\left[(1+\delta)^{3/2}e^{-2m|1+\delta|r\sin(\pi/q)}\right.\nonumber\\
	&+&\left.(1-\delta)^{3/2}e^{-2m|1-\delta|r\sin(\pi/q)} \right] \  . 
\end{eqnarray}
For $1\leq q < 2$ the sum over $k$ in \eqref{T00z} is absent
and the integral terms are suppressed by the factor $e^{-2|1\pm \delta|mr}$. 

\begin{figure}[h]
	\centering
	{\includegraphics[width=0.48\textwidth]{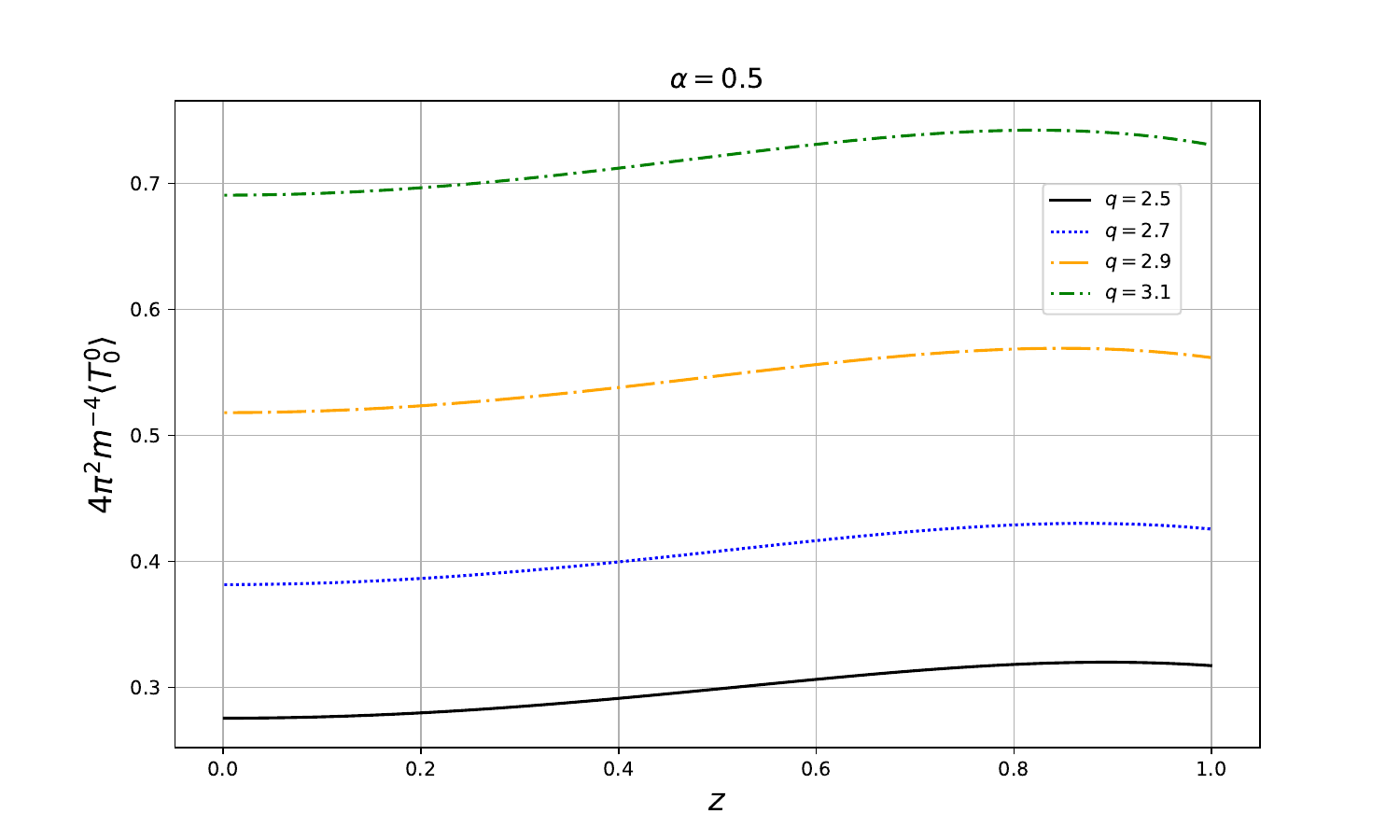}}
	\hfill
	{\includegraphics[width=0.48\textwidth]{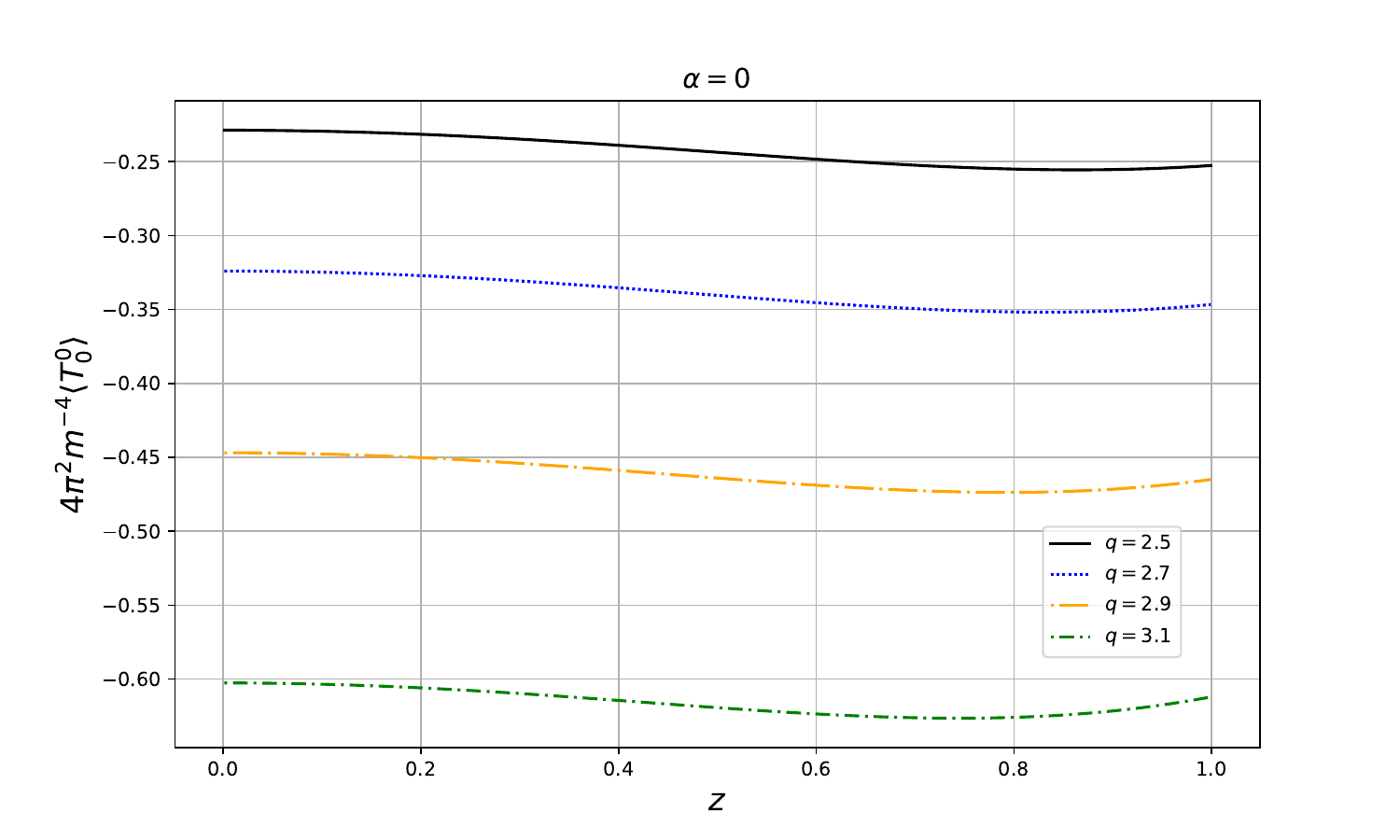}}
	\caption{The quantity $4\pi^2\langle T_0^0 \rangle ^{\mathrm{ren}}/m^4$, is exhibited as 
		function of $\delta$ for $\alpha=1/2$  (left plot) and $\alpha=0$ (right plot)
		for  $q=2.5, \ 2.7, \ 2.9, \ 3.1  $.  In both plots we consider
		$\delta$ in the interval $[0, \ 1]$ and assume $mr=1.0$.}
	\label{fig03}
\end{figure}

In Fig. \ref{fig03}  we exhibit the behavior of the energy-density  as function of $\delta$ 
for $\alpha=1/2$ (left plot) and $\alpha=0$ (right plot),  considering different values of $q$.
In these plots we take $mr=1$.  Moreover, we can observe that the intensities of the module of the 
energy-densities increase with $q$.

\begin{figure}[h]
	\centering
	{\includegraphics[width=0.48\textwidth]{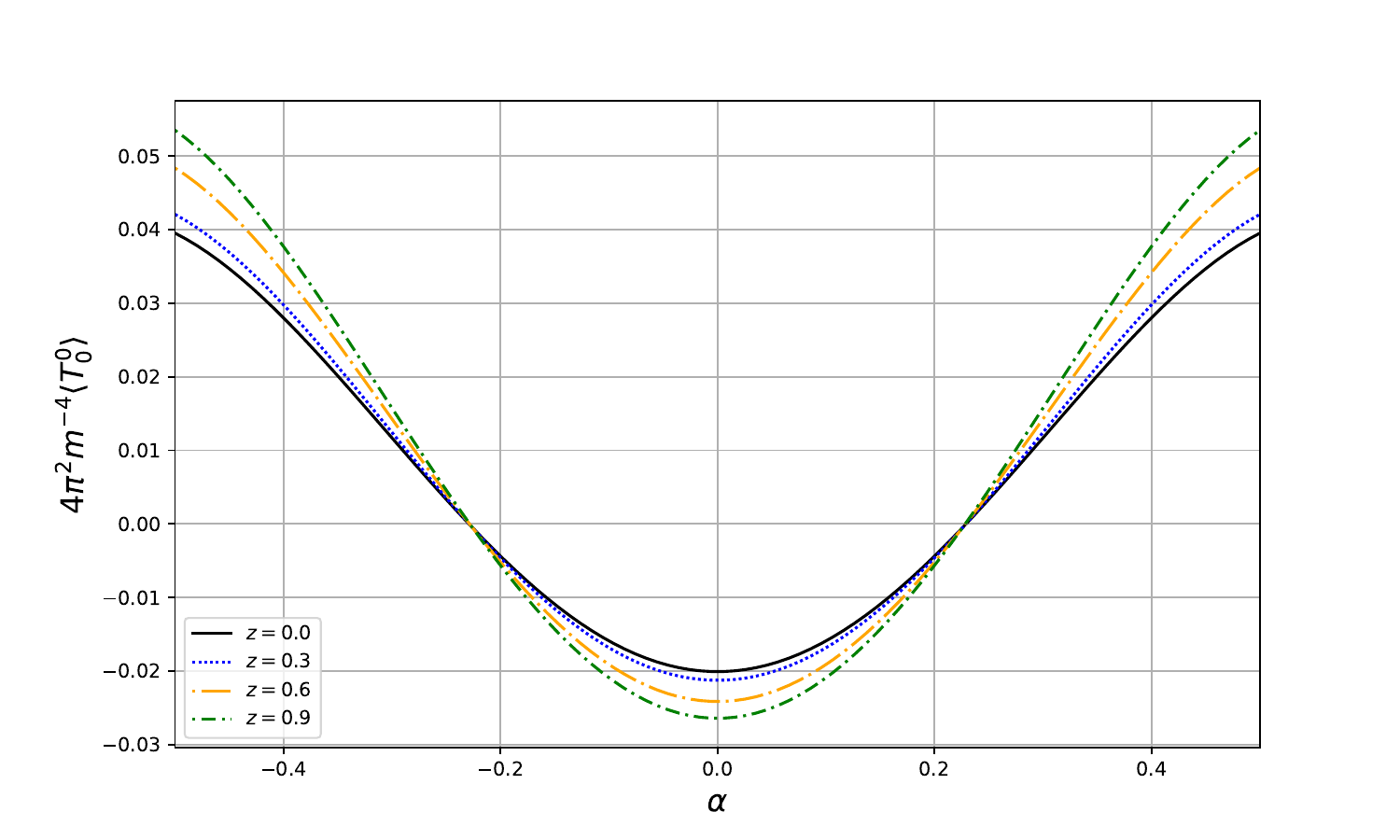}}
	\caption{The energy-density, $4\pi^2\langle T_0^0 \rangle ^{\mathrm{ren}}/m^4$,
		is exhibited as function of $\alpha$ for  
		$\delta=0.0, \ 0.3, \  0.6, \ 0.9 $. In this plot was considered 
		$mr=1$ and $q=1.5$.}
	\label{fig04}
\end{figure}

Figure \ref{fig04} presents the dependence of the energy-density as function of $\alpha$ 
for different values of $\delta$, considering $mr=1$ and $q=1.5$. By this graph we can see 
that the energy density can be positive or negative, depending on the value of $\alpha$,
for every value of $\delta$ considered. By this plot we can see that the energy-density, $\langle T_0^0\rangle^{ren}$,
can assume positive or negative values. For vanishing magnetic flux, $\alpha=0$, this quantity is 
always negative, by turning on the magnetic flux, the corresponding Aharonov-Bohm effect
provides a growth in the intensity of this observable. 

\section{Conclusion}
\label{conc}

The analysis of fermionic and bosonic vacuum polarizations induced by 
an Abelian vortex configuration taking into account the conical
geometry of the spacetime produced by this topological defect, 
have been developed by many researchers, great part of them cited in 
this paper; however, for the non-Abelian vortex configuration, as fas as
we know, this analysis has been missed. Here in this paper we try to 
fill that gap. In fact the analysis developed in this paper is for the
$SU(2)$ vortex field. For this specific configuration, 
only iso-vectors field can be considered. In this way we 
consider the system of isopin$-1/2$ fermionic field coupled with
iso-scalars and iso-vector gauge fields. Because the
complete analysis of this system is almost impossible to be developed,
we assume two important simplifications: $i)$ that the vortex 
is an idealized topological defect, i.e., very thin and straight. So 
we can discard the inner structure of the iso-vectors, gauge and scalars field, and
$ii)$ the second is to consider the coupling of the fermion with 
iso-scalar $\chi^a$, given in \eqref{chi_1}. So, the $8\times8$ matrix
Hamiltonian could be written in a diagonal form, Eq. \eqref{Repres}. 
As consequence the $8-$component isopin$-1/2$ fermionic field 
effectively behaves as two-independent $4-$component ones.
Each of them presenting different energy as exhibit in \eqref{Energy_function}
and different normalizations constants. The corresponding fermionic modes,
 $\psi^{(\pm)}_{\sigma(+)}$ and $\psi^{(\pm)}_{\sigma(+)}$, are
given in \eqref{Solution_1} and \eqref{Solution_2}, expressed 
in terms of Bessel functions with  orders \eqref{betaj} and
\eqref{betaj1}, respectively. 

Having obtained the complete set of normalized isospin$-1/2$ fermionic modes,
we proceed to the calculations of the FC and the VEV of the energy-momentum tensor.
Due to the presence of the coupling with the iso-scalar field, there appears
different corrections in the mass term for the fermionic Hamiltonian, 
$H_{(+)}$ and $H_{(-)}$, given in 
\eqref{Hamilt0}.  In the calculation of the FC by equation \eqref{FC1}, 
we changed in the summation over the eigenvalues of the total angular momentum
$j$ by $-j$ in the second contribution, providing the expression \eqref{FC2}.
The next steps was to obtain the renormalized FC. After some intermediate steps,
our final result was given by Eq. \eqref{FC4}.
Defining by $\delta$ the ratio $g_Y/m$, the renormalized FC could be expressed by 
\eqref{FC5}. This FC presents two distinct contributions: one due to the conical topology
of the background spacetime, and the other due to the interaction of the
fermionic field with the magnetic flux running along the vortex.
From our result we can observe that for massless field the FC vanishes. 
Some asymptotic behavior of this quantity were developed:  for large distance from
the string; it decays as $e^{-2|1\pm \delta|mr\sin(\pi/q)}$ for $q\geq 2$, and as $e^{-2|1\pm \delta|mr}$
for $1\leq q < 2$. Near the string the FC behaves as $1/r^2$, as exhibited in \eqref{FC5m}.
In Eq. \eqref{FC6}, we provide the asymptotic behavior
for fixed distance and $\delta\to\pm1$. Because the analysis of the behavior of the FC 
with the parameter $\delta$, cannot be observed in terms of analytical functions, 
we only can  answer this question numerically.
In Fig. \ref{fig01} we exhibit the behavior of the 
FC as function of $\delta$ in the interval $[0, \ 1]$, for $\alpha=1/2$ (left plot), and 
$\alpha=0$ (right plot), for different values of $q$. By these plots we conclude that 
the module of the intensity increases with $q$ for a given $\delta$. On the
other hand, for a given $q$ the modulus of the intensity is higher for
$\delta=0$, that corresponding the absence of coupling between the
fermionic field with the bosonic sector of the system.  In Fig. \ref{fig02}
we exhibit the behavior of the FC with $\alpha$ for different values of $\delta$ keeping 
$q=3.5$ and $mr=1$. We note that FC is higher for $\delta=\alpha=0$. In fact the
interaction of the fermionic field with the magnetic flux, provides a decreases in the FC; 
moreover, for a given $\alpha$, the coupling with the iso-scalar field contributes to the 
imbalance between the contributions of both
effective masses, decreasing, by its turn, the magnitude of FC. 
In addition we can see that the FC can be positive or negative depending the value 
of $\alpha$, and it seems that this changing in the signal occurs for all $\delta$ at the
same value.

Another important characteristic of the fermionic vacuum is the VEV of the 
energy momentum tensor.  In section \ref{EMT0} we have presented the mains procedures 
concerning with all calculations.  Specifically the obtainment of the
renormalized energy-density required  many steps given in detail in the corresponding
subsection; one of them was in the changing of the quantum number $j\to-j$
in the summation over the contribution 
associated with the mode $\psi_{\sigma(-)}^{(\pm)}$. As to the other
components, they could be expressed in terms of $\langle T^0_0\rangle^{ren}$. The 
renormalized expression for energy-density was presented in \eqref{T003}. 
The expression for $\langle T^0_0\rangle^{ren}$, considering
massless field, but $g_Y\neq 0$ is given in \eqref{T003_m}.
Finally considering in the latter the limit $|g_y|r\to0$ the energy density
behaves as $1/r^4$, as shown in \eqref{T003_m0}.
The two more important properties satisfied by the VEV of the energy-momentum 
tensor, have been presented in section \ref{sec5}. They are:
trace relation and the conservation condition. 
In addition some important analytical
result were obtained for some limiting values of the
physical parameter of the theory.
For large distance from the string, and for $q\geq 2$, 
the energy-density decays as $e^{-2m|1\pm \delta|r\sin(\pi/q)}$,
and decays with $e^{-2|1\pm \delta|mr}$ for $1\leq q <2$. 
Similarly what we have pointed out in the last section, 
the knowledge of the dependence of the energy-density with the parameters
$\delta$ and $q$, can oly be provided numerically. 
With this objective, in Fig \ref{fig03} we exhibit the behavior of $\langle T^0_0\rangle^{ren}$
as function of $\delta$ defined in the interval $[0, \ 1]$ for $\alpha=1/2$ (left panel) and
$\alpha=0$ (right panel), considering different values of $q$.
The Fig. \ref{fig04} exhibits its behavior as function of
$\alpha$, taking $mr$ and $q=1.5$, and varying $\delta$. We note
that the energy-density  can be positive or negative depending on the value of
$\alpha$ for each value of $\delta$. 

 At this point we have to say that we have calculated 1-loop energy-momentum tensor 
generated by the fermionic field. Another source
of contributions are due to the bosonic fields. Because we 
are considering that the energy scale where the gauge symmetry is broken is much larger than
the energy associated with the fermionic field, it is expected that 
the contributions of the Higgs and vector fields
are suppressed by some negative power of their corresponding masses. 
However considering only the gauge and matter fields such contributions are important since they 
allow to obtain analytic results for the vacuum energy \cite{Vassievich,Rebhan}. 
In addition, the analysis of stability of the $SU(2)$ Higgs-gauge system taking into account
quantum effects has been developed in \cite{Graham}. Finally a review about the  computation
of 1-loop order corrections to the classical masses of some topological object,
specifically the planar Abelian-Higgs model, with respective Feymann rules, 
have been provided in \cite{Alonson}.
 
Finally, as our last comments, we would like to say that the singular behavior of the FC and the energy-density
near the vortex, is consequence of the idealized model adopted for this topological
defect. In this way, we considered for the vortex and
iso-scalar fields their vacuum values.
Unfortunately there is no analytical function capable to provide the behavior associated 
with the non-Abelian vortex field for all values of radial distance $r$, only numerical representation can do that. 
Approximated model may be adopted, like in the case of Abelian vortex \cite{Bordag93,Bordag94}.
This procedure deserves to be developed in a separate work, as the natural continuation. In this sense we can
say that the present calculation can be considered as the first step in the
analysis of fermionic vacuum polarization induced by the presence of non-Abelian vortex. Still considering
the idealized model for the non-Abelian vortex, a possible future calculation
could be the obtainment of the induced fermionic current.

\section*{Acknowledgments}

E.R.B.M is partially supported by CNPq under Grant no 301.783/2019-3.
H.F.S.M is partially supported by CNPq under Grants no. 305379/2017-8 and 
311031//2020-0.


\begin{thebibliography}{99}

\bibitem{Kibble1976} T. W. B. Kibble, J. Phys. A {\bf 9}, 1387 (1976).

 \bibitem{Vilenkin-Shellard} A. Vilenkin and E. S. Shellard, {\it Cosmic 
 strings and others topological defects}.  Cambridge University Press (2000).

\bibitem{N-O} H. B. Nielsen and P. Olesen, Nuclear Physics B {\bf61}, 45 (1973).

 \bibitem{DG} D. Garfinkle, Phys. Rev. D {\bf 31} 1323 (1985).

 \bibitem{Laguna} P. Laguna-Castillo and R. A. Matzner, Phys. Rev. D {\bf 35} 2933 (1987).

\bibitem{Linet1} B. Linet, Phys. Lett. B \textbf{124}, 240 (1987).

\bibitem{Padua_2015} A. de P\'adua Santos and E. R. Bezerra de Mello, Class.
Quantum Grav. {\bf 32} 155001 (2015).

\bibitem{Hell86} T.M. Helliwell and D.A. Konkowski, Phys. Rev. D \textbf{34}
, 1918 (1986).

\bibitem{Line87} B. Linet, Phys. Rev. D \textbf{35}, 536 (1987).

\bibitem{Frol87} V.P. Frolov and E.M. Serebriany, Phys. Rev. D \textbf{35},
3779 (1987).

\bibitem{Dowk87} J.S. Dowker, Phys. Rev. D \textbf{36}, 3095 (1987); J.S.
Dowker, Phys. Rev. D \textbf{36}, 3742 (1987).

\bibitem{Davi88} P.C.W. Davies and V. Sahni, Class. Quantum Grav. \textbf{5}
, 1 (1988).

\bibitem{Smit89} A.G. Smith, in \textit{The Formation and Evolution of
Cosmic Strings}, Proceedings of the Cambridge Workshop, Cambridge, England,
1989, edited by G.W. Gibbons, S.W. Hawking, and T. Vachaspati (Cambridge
University Press, Cambridge, England, 1990).

\bibitem{Alle90} B. Allen and A.C. Ottewill, Phys. Rev. D \textbf{42}, 2669
(1990); B. Allen, J.G. Mc Laughlin, and A.C. Ottewill, Phys. Rev. D \textbf{%
45}, 4486 (1992); B. Allen, B.S. Kay, and A.C. Ottewill, Phys. Rev. D
\textbf{53}, 6829 (1996).

\bibitem{Sour92} T. Souradeep and V. Sahni, Phys. Rev. D \textbf{46}, 1616
(1992).

\bibitem{Shir92} K. Shiraishi and S. Hirenzaki, Class. Quantum Grav. \textbf{%
9}, 2277 (1992).

\bibitem{Beze94} V.B. Bezerra and E.R. Bezerra de Mello, Class. Quantum
Grav. \textbf{11}, 457 (1994); E.R. Bezerra de Mello, Class. Quantum Grav.
\textbf{11}, 1415 (1994).

\bibitem{Cogn94} G. Cognola, K. Kirsten, and L. Vanzo, Phys. Rev. D \textbf{%
49}, 1029 (1994).

\bibitem{More95} E.S. Moreira Jnr, Nucl. Phys. B \textbf{451}, 365 (1995).

\bibitem{Iell97} D. Iellici, Class. Quantum Grav. \textbf{14}, 3287 (1997).

\bibitem{Khus99} N.R. Khusnutdinov and M. Bordag, Phys. Rev. D \textbf{59},
064017 (1999).

\bibitem{BezeKh06} V.B. Bezerra and N.R. Khusnutdinov, Class. Quantum Grav.
\textbf{23}, 3449 (2006).

\bibitem{charged} J.S. Dowker, Phys. Rev. D \textbf{36}, 3742 (1987).

\bibitem{charged1} M.E.X. Guimar\~{a}es and B. Linet, Commun. Math. Phys.
\textbf{165}, 297 (1994).

\bibitem{charged3} J. Spinelly and E.R. Bezerra de Mello, Class. Quantum
Grav. \textbf{20} 874, (2003); J. Spinelly and E.R. Bezerra de Mello, Int.
J. Mod. Phys. A, \textbf{17}, 4375 (2002).

\bibitem{Spin} J. Spinelly and E.R. Bezerra de Mello, Int. J. Mod. Phys. D
\textbf{13}, 607 (2004); J. Spinelly and E.R. Bezerra de Mello, Nucl Phys. B
(Proc. Suppl.) \textbf{127}, 77 (2004).

\bibitem{Spin08} J. Spinelly and E. R. Bezerra de Mello, JHEP \textbf{09},
005 (2008).

\bibitem{Vega78} H. J. de Vega, Phys. Rev. D {\bf 18} 2932 (1987).

\bibitem{Mello87} E. R. Bezerra de Mello, J. Phys. G: Nucl. Phys. {\bf 13} 897 (1987).

\bibitem{Mello88a} E. R. Bezerra de Mello, Phys. Rev. D {\bf 38} 2616 (1988).

\bibitem{Mello88b}  E. R. Bezerra de Mello, Phys. Rev. D {\bf 38} 3755  (1988).

\bibitem{Jackiw_1981} R. Jackiw and P. Rossi, Nucl Phys. B {\bf 190} 681 (1981)

\bibitem{Saha13} E. R. Bezerra de Mello and A. A. Saharian, Eur. Phys. J. C.

\textbf{73}, 2532 (2013).

\bibitem{Saha14} S. Bellucci,  E. R. Bezerra de Mello, A de Padua and A. A. Saharian, Eur. Phys. J. C.
\textbf{74}, 2688 (2014).

\bibitem{Vega_86} H. J. de Vega and F. A. Schaposnik, Phys. Rev. Lett, {\bf 56}, 2564 (1986).

\bibitem{Rubakov} V. A. Rubakov, Nuc. Phys. B {\bf 203}, 311 (1982).

\bibitem{Greiner} W. Greiner, {\it Quantum Mechanics: an introduction} (4th Edition, Spring-Verlag  
Berlin Heidelberg, 2001.)


\bibitem{Klim88} K.G. Klimenko, Phase structure of generalized Gross-Neveu
models. Z. Phys. C \textbf{37}, 457 (1988).

\bibitem{Eliz94} E. Elizalde, S. Leseduarte, S.D. Odintsov, Chiral symmetry
breaking in the Nambu-Jona-Lasinio model in curved spacetime with a
nontrivial topology. Phys. Rev. D \textbf{49}, 5551 (1994).

\bibitem{Grad} I.S. Gradshteyn and I.M. Ryzhik, \textit{Table of Integrals,
Series and Products} (Academic Press, New York, 1980).

\bibitem{Saha10} E. R. Bezerra de Mello, V.B. Bezerra, A.A. Saharian, and
V.M. Bardeghyan, Phys. Rev. D \textbf{82}, 085033 (2010).

\bibitem{Mello13} E.R. Bezerra de Mell, A.A. Saharian, and S.V. Abajyan,
Class. Quantum Grav. \textbf{30}, 015002 (2013).


\bibitem{Abra} \textit{Handbook of Mathematical Functions}, edited by M.
Abramowitz, I.A. Stegun (Dover, New York, 1972).

\bibitem{Bordag93} M. Bordag and S. Voropaev, J. Phys. A: Math. Gen. {\bf 26}, 7637 (1993).

\bibitem{Bordag94} M. Bordag and S. Voropaev, Phys. Lett. B {\bf 333}, 238 (1994).

\bibitem{Vassievich} D. V. Vassilevich, Phys. Rev. D {\bf 68},  045005 (2003).

\bibitem{Rebhan} A. Rebhan, P. van Nieuwenhuizen and R. Wimmer, Nucl. Phys. B {\bf 679},  382 (2004).

\bibitem{Graham} N. Graham and H. Weigel, Int. J. Mod. Phys. A {\bf 37}, 2241004 (2022).

\bibitem{Alonson} A. Alonso Izquierdo, W. Garcia Fuertes, M.A. Gonzalez Leon
M. de la Torre Mayado, J. Mateos Guilarte and J.M. Mu\~noz Casta\~neda, 
{\it Heat kernel/Zeta function control of one-loop divergences.} hep-th/0611180.

\end{thebibliography}
\end{document}